\documentclass[aps,prb,twocolumn,showpacs,amsmath,amssymb,floatfix]{revtex4-1}

\newcommand{\be}{\begin{equation}}
\newcommand{\ee}{\end{equation}}
\newcommand{\bea}{\begin{eqnarray}}
\newcommand{\eea}{\end{eqnarray}}

\usepackage{graphicx}
\usepackage{dcolumn}
\usepackage{bm}
\usepackage{subfigure}
\usepackage{color}

\begin{document}

\title{Persistent charge and spin currents in a 1D ring with Rashba and Dresselhaus
spin-orbit interactions by excitation with a terahertz pulse}

\author{Marian Ni\c t\u a}
\affiliation{National Institute of Materials Physics, P.O. Box MG-7,
Bucharest-Magurele, Romania}

\author{D.\ C.\ Marinescu}
\affiliation{Department of Physics and Astronomy, Clemson University, Clemson,
South Carolina 29634, USA}

\author{Andrei Manolescu}
\affiliation{School of Science and Engineering, Reykjavik University, Menntavegur 1,
IS-101 Reykjavik, Iceland}

\author{Bogdan Ostahie}
\affiliation{National Institute of Materials Physics, P.O. Box MG-7,
Bucharest-Magurele, Romania}

\author{Vidar Gudmundsson}
\affiliation{Science Institute, University of Iceland, Dunhaga 3, IS-107 Reykjavik,
Iceland}

\begin{abstract}
Persistent, oscillatory charge and spin currents are shown to be driven
by a two-component terahertz laser pulse in a one-dimensional mesoscopic
ring with Rashba-Dresselhaus spin orbit interactions (SOI) linear in the
electron momentum. The characteristic interference effects result from
the opposite precession directions imposed on the electron spin by the two
SOI couplings. The time dependence of the currents is obtained by solving
numerically the equation of motion for the density operator, which is
later employed in calculating statistical averages of quantum operators
on few electron eigenstates. The parameterization of the problem is done
in terms of the SOI coupling constants and of the phase
difference between the two laser components.  Our results indicate that
the amplitude of the oscillations is controlled by the relative strength
of the two SOI's, while their frequency is determined by the difference
between the excitation energies of the electron states. Furthermore,
the oscillations of the spin current acquire a beating pattern of higher
frequency that we associate with the nutation of the electron spin between
the quantization axes of the two SOI couplings. This phenomenon disappears
at equal SOI strengths, whereby the opposite precessions occur with the
same probability.
\end{abstract}


\maketitle

\section{Introduction}

The reconsideration of the spin-orbit interaction (SOI) in the modern
context of spintronic applications, started more than ten years ago, was
based on the idea of the possible manipulation of the electron spin by
means of an electric field.\cite{datta} While definitive answers to this
quest have yet to be reached, the fundamental interest in understanding
the effect of SOI on macroscopic phenomenology continues. Associated with
the spatial confinement in two-dimensional quantum wells\cite{rashba}
(Rashba-R) or with inversion asymmetry in the crystal (Dresselhaus-D)
\cite{dresselhaus}, SOI is usually considered to be linear in the
electron momentum which is coupled to the electron spin through a
constant whose magnitude is amenable to outside control. The most
important physical aspect associated with the R-D superposition
is the effect on the electron spin, which is forced to precess in
opposite directions along the quantization directions imposed by the
two spin interactions. This phenomenology is embodied by the single particle
Hamiltonian,
\begin{equation}
H_{SO}=\frac{\alpha}{\hbar}(\sigma_x p_y-\sigma_y p_x) +
\frac{\beta}{\hbar}(\sigma_x p_x-\sigma_y p_y) \ ,
\label{eq:ho}
\end{equation}
which introduces the Rashba and Dresselhaus interaction constants $\alpha$
and $\beta$ which couple the electron momentum $\vec{p}=(p_x,p_y)$ with
the electron spin represented by the Pauli spin matrices, $\sigma_{x,y}$.
The superposition of the Rashba and Dresselhaus terms generates
particularly interesting situations when their strengths are equal,
such as the cancellation of dephasing for the eigenstate spinors and
the ensuing ballistic spin transport\cite{egues} or the formation of
a persistent spin helix.\cite{bernevig}

In this paper, we investigate the consequences of the  Rashba-Dresselhaus
superposition on spin and charge currents that are being induced in
a quasi-one-dimensional ring by a terahertz laser pulse with
a spatial asymmetry. This represents a well known method of current
generation which exploits the left-right asymmetry of the electron states
that are excited on a time scale that is much shorter than the electron
relaxation lifetime in a non-adiabatic process.\cite{gudmundsson, siga}
In general, by this method, as well as by applying magnetic fields,
\cite{moskalets1,moskalets2} simultaneous spin and charge current
generation occurs.\cite{splettstoesser, souma, sheng} The independent
realization of pure spin currents has been addressed in a number of
theoretical proposals that considered hybrid structures\cite{sun} or a
specific electron configuration (odd numbers of particles).\cite{huang}
More recently it was shown that a pure spin current can be created
non-adiabatically in a ring with Rashba interaction using a
radiation pulse with two dipolar components having a spatial dephasing
angle $\phi$.\cite{nita1} The physical mechanism for spin current
generation relies on the interplay between the spin orbit coupling that
rotates the electron spin around the ring and the spatial asymmetry of
the external excitation, which establish conditions where the
charge current disappears, while the spin current reaches a maximum or
a minimum level.

This work undertakes the analysis of spin and charge dynamics in
the simultaneous presence of Rashba and Dresselhaus SOI, such that
two opposite precession directions are imposed on the electron spin. Our
numerical results are obtained within an equation-of-motion algorithm for
the particle-density operator, that is later involved in calculating the
spin and charge currents as statistical averages of the corresponding
quantum operators on a few non-interacting electron eigenstates.  The
internal phase difference $\phi$ between the two laser components along
with the coupling strengths of the R and D interactions are the main
parameters of the problem. While the latter play a role in determining
the magnitude of the effects, the former is shown to influence
the spatial distribution of the currents around the ring.

The paper starts by discussing the spectrum of the equilibrium
Hamiltonian, in Sec.~\ref{sec:eq-ham}, whose eigenstates and eigenvalues
are obtained within a direct diagonalization procedure for a small
number of electrons. Then, in Sec.~\ref{sec:dens-op} we detail the
equation-of-motion algorithm that allows the estimation of the density
operator, followed by Sec.~\ref{sec:sym} and \ref{sec:asym} that present
our findings. The results are summarized in Sec.~\ref{sec:conc}.

\section{The Equilibrium Hamiltonian}\label{sec:eq-ham}

Our analysis is based on a discrete model of the Hamiltonian that relies
on transforming the continuous, quasi-one dimensional ring of radius
$r_0$ into a sequence of $N$ sites (points) separated by an equivalent
lattice constant $a=2\pi r_0/N$.  In polar coordinates a site of index
$n=1,...,N$ has an azimuthal angle $\theta_n=2\pi n/N$, the angle difference
between two consecutive sites being $\Delta\theta=2\pi/N$. A single particle
state that corresponds to an electron of spin $\sigma$ located at point
$n$, $|n,\sigma\rangle$, is associated with the creation
and annihilation operators $c^{+}_{n,\sigma}$ and $c_{n,\sigma}$. The
electrons encased in the ring are described by the
Hamiltonian $H_{\rm ring}$ composed out of the kinetic energy $H_0$
and the Rashba and Dresselhaus terms, $V_R$ and $V_D$,
\be
H_{\rm ring}=H_0+V_{R}+V_{D} \,,
\label{h}
\ee
where $V_R$ and $V_D$ are given by the first and the second term
of Eq.\ (\ref{eq:ho}), respectively. In the representation provided by
the single particle states described above, $H_0$ is proportional to
the hopping energy $V=\hbar^2/2m^*a^2$ ($m^*$ being the effective mass of
the electron in the host semiconductor material),
\be
H_0=2V\sum_{n,\sigma}c^{+}_{n,\sigma}c^{}_{n,\sigma}
-V\sum_{n,\sigma} c^{+}_{n,\sigma}c^{}_{n+1,\sigma}
-V\sum_{n,\sigma} c^{+}_{n,\sigma}c^{}_{n-1,\sigma} \ ,
\label{h0}
\ee
while the SOI terms $V_R$ and $V_D$ generate their own energy scales,
$V_{\alpha}=\alpha/2a$ and $V_{\beta}=\beta/2a$,
\be\label{vr}
      V_{R}=
      -iV_{\alpha}\sum_{n,\sigma,\sigma'}
      \left[ {\bf \sigma}_r(\theta_{n,n+1}) \right]_{\sigma, \sigma'}
      c^{+}_{n \sigma} c^{}_{n+1 \sigma'}+{\rm H.c.}\, ,
\ee
and
\be\label{vd}
      V_{D}=
      -iV_{\beta}\sum_{n,\sigma,\sigma'}
      \left[ {\bf \sigma}_\theta(\theta_{n,n+1}) \right]^*_{\sigma, \sigma'}
      c^{+}_{n \sigma} c^{}_{n+1 \sigma'}+{\rm H.c.} \, .
\ee
For simplicity, it is customary to introduce the azimuthal and radial spin
matrices $\sigma_\theta$ and $\sigma_r$ written in terms of the angle
$\theta_{n,n+1}=(\theta_n +\theta_{n+1})/2$,
\begin{eqnarray}
{\bf \sigma}_r(\theta)&&={\bf \sigma}_x \cos\theta+{\bf \sigma}_y \sin\theta \, , \nonumber \\
{\bf \sigma}_\theta(\theta)&&=-{\bf\sigma}_x \sin\theta+ {\bf \sigma}_y \cos\theta\, .
\end{eqnarray}
In the presence of only one type of SOI, say Rashba, the spectrum of
the Hamiltonian, as well as the eigenvalues  of the spin operator, can
be obtained analytically.  The single-particle eigenvectors are also
eigenstates of $L_z$, the $\hat{z}$ component of the angular momentum,
and they are
\begin{eqnarray}\label{eigenvectorsR}
&&|\Psi_{l+}\rangle=\frac{1}{\sqrt{N}}\sum_n e^{il\theta_n}
\left( \begin{array}{c} \cos \theta_\alpha\\ - e^{i\theta_n}\sin \theta_\alpha
\end{array}
\right) |n\rangle \,  \nonumber \\
&&|\Psi_{l-}\rangle=\frac{1}{\sqrt{N}}\sum_n e^{il\theta_n}
\left( \begin{array}{c}
e^{-i\theta_n}\sin \theta_\alpha  \\
\cos \theta_\alpha
\end{array}
\right) |n\rangle \, ,
\end{eqnarray}
with $l=0,\pm 1,\cdots , \pm (N-1)/2, N/2$ (for N even).
Correspondingly, the energy eigenvalues are given by
\begin{eqnarray}\label{eigen1}
E_{l\pm}=\frac{\epsilon_l+\epsilon_{l\pm 1}}{2} +
\frac{\epsilon_l- \epsilon_{l\pm 1}}{2}\sqrt{1+\tan^2 2\theta_\alpha} \, ,
\end{eqnarray}
where $\epsilon_l=2V-2V\cos(2\pi l/N) \,$  and 
$\tan 2\theta_\alpha=V_\alpha/({V\sin(\pi/N)})$.  The quantization direction
of the spin operator is ${\bf e}_{2\theta_\alpha }=\cos 2\theta_\alpha
{\bf e}_z -\sin 2\theta_\alpha {\bf e}_r$, which is tilted at an angle
$2\theta_{\alpha}$ relatively to the $\hat{z}$-axis due to the presence of
the of the SOI.\cite{splettstoesser} The eigenstates are spin degenerated
and the energies $E_{l\pm}$ correspond to spin eigenvalues of $\pm
\hbar/2$ along ${\bf e}_{2\theta_\alpha }$.  In turn, the Dresselhaus
interaction alone determines a similar energy spectrum and eigenstates,
but defines another preferential spin orientation direction, ${\bf
e}_{2\theta_\beta}=\cos 2\theta_\beta {\bf e}_z +\sin 2\theta_\beta
{\bf e}_\theta^*$ where ${\bf e} _\theta^*=-{\bf e}_{-\theta}$ (${\bf
e}_{-\theta}$ is azimuthal direction for the point of angle $-\theta$
of the ring).

\begin{figure}
\centering
\includegraphics[width=70mm]{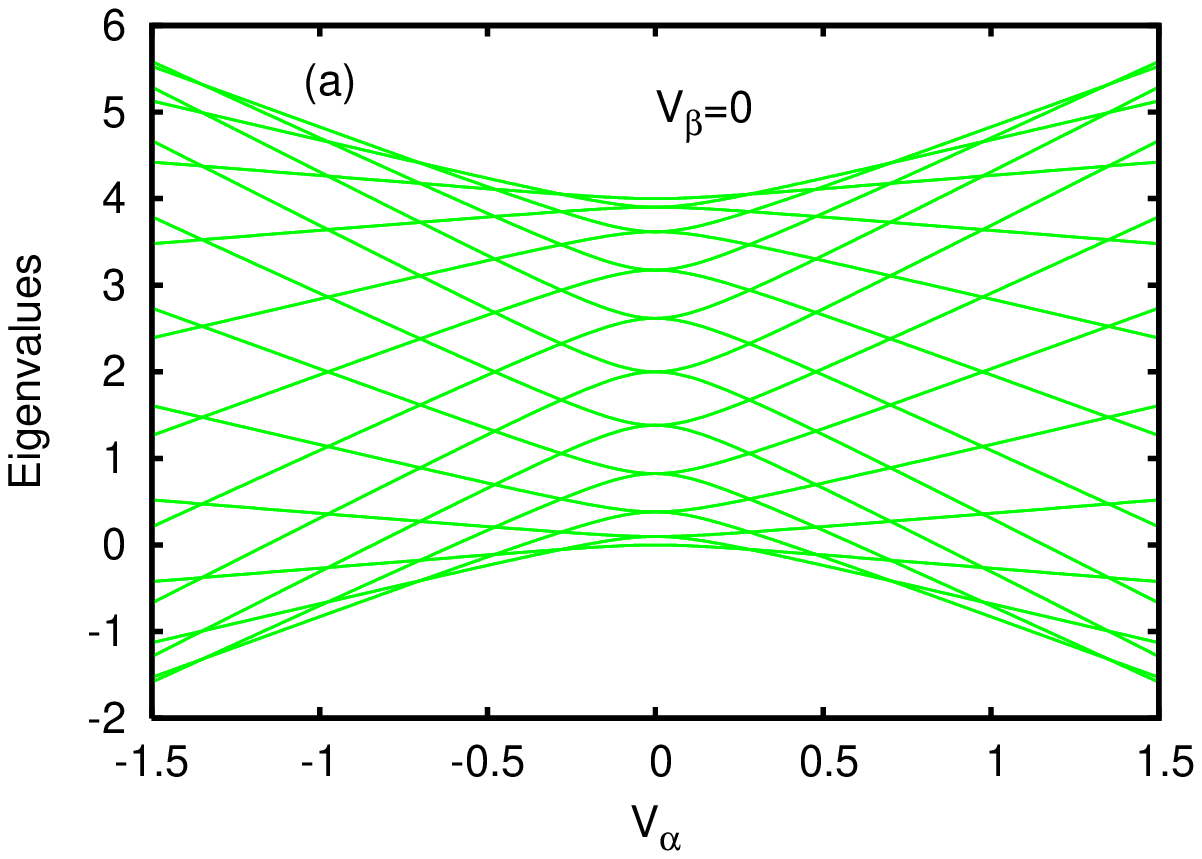}
\includegraphics[width=70mm]{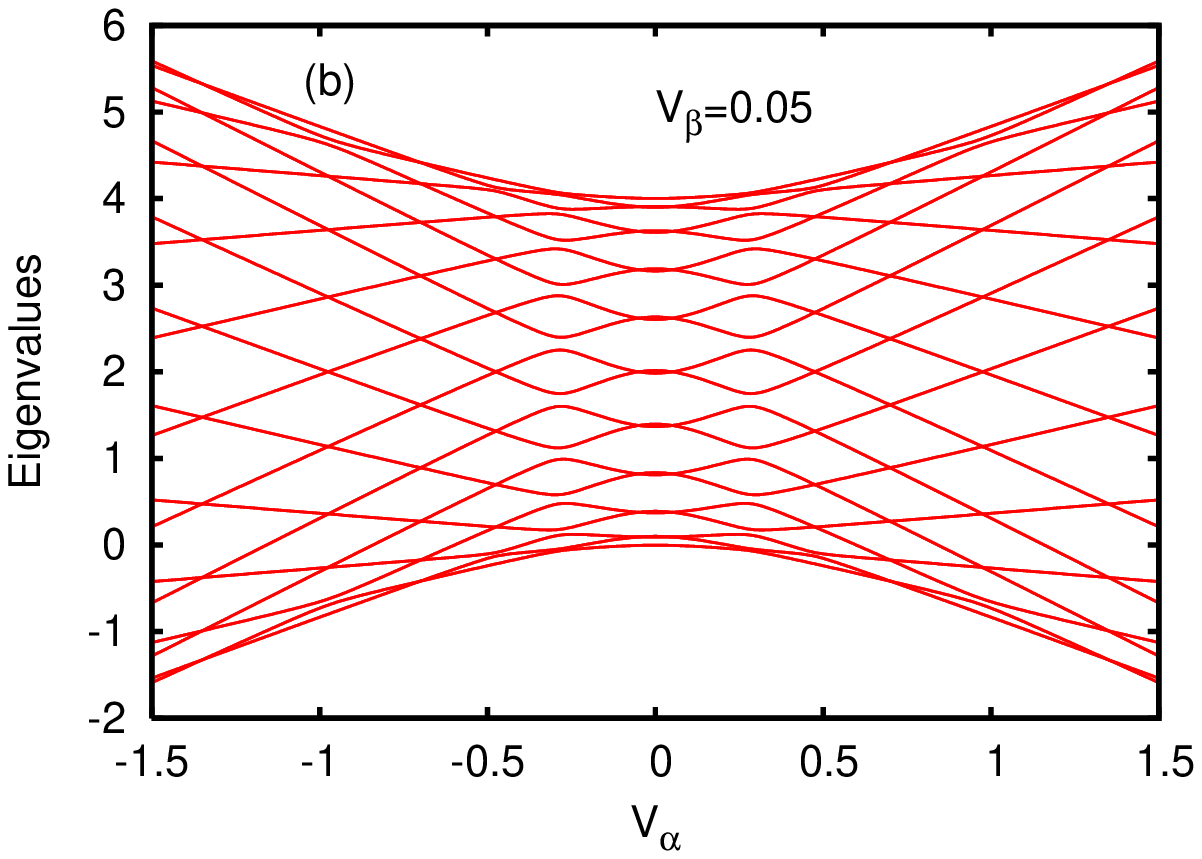}
\caption{(Color online) 
The energy spectrum of a ring with SOI versus Rashba energy $V_\alpha$.
The Dresselhaus energy is $V_\beta = 0$ in (a) and $V_{\beta}=0.05$ in (b).
All energies are given in  units of $V=\hbar^2/2m^*a^2$.  The spectrum has 
40 eigenstates, corresponding to 20 sites $\times$ 2 spin directions.  
All eigenstates are spin-dengenerated.
}
\label{spectre}
\end{figure}

When both R and D interactions are present, the diagonalization
of the Hamiltonian (\ref{h}) is possible only by numerical methods
and the results are strongly affected by the relative values of the
two coupling constants $\alpha$ and $\beta$.  The energy spectrum of
the equilibrium Hamiltonian, calculated for an even number of sites,
$N = 20$, is shown in Fig.~\ref{spectre} as a function of $V_\alpha$,
for different values of $V_\beta$.
All states remain spin degenerate, but other degeneracies are lifted.
When $V_\beta=0$, as in Fig.\,\ref{spectre}(a), the spectrum is
four fold degenerate for those values of $V_\alpha$ that allow the
equality $E_{l+}=E_{l'-}$ for two different quantum numbers $l$
and $l'$. Such degeneracies are lifted when $V_\beta\ne 0$
and comparable to $V_\alpha$, as shown in Fig.\,\ref{spectre}(b),
where we notice that small energy gaps appear between previously
intersecting levels.  These gaps give rise to low frequency oscillations
of spin and charge currents.

The azimuthal velocity operator $v_\theta=r_0d\theta/dt = (ir_0/\hbar)[H,\theta]$
is evaluated from Eq.\ (\ref{h}),
\begin{eqnarray}\label{vrd}
&&v_\theta=-i\frac{Va}{\hbar}\sum_{n,\sigma}[ c_{n,\sigma}^{\dagger}c_{n+1,\sigma}^{}-c_{n,\sigma}^{\dagger}c_{n-1,\sigma}^{}]\nonumber\\
&&+2\frac{V_\alpha a}{\hbar}\sum_{n,\sigma,\sigma'}{\bf \sigma}_r(\theta_n)^{\sigma,\sigma'} c_{n,\sigma}^{\dagger}c_{n,\sigma'} \nonumber \\
&&+2\frac{V_\beta a}{\hbar}\sum_{n,\sigma,\sigma'}{\bf \sigma}^*_\theta(\theta_n)^{\sigma,\sigma'} c_{n,\sigma}^{\dagger}c_{n,\sigma'},
\end{eqnarray}
leading to the charge and spin current operators:
\begin{eqnarray}\label{jz}
I & = & -e v_\theta \;, \nonumber\\
I^s_{\nu}& =& \frac{\hbar}{2} \left( {\bf \sigma}_\nu v_\theta
                        +v_\theta {\bf \sigma}_\nu \right) \; .\label{eq:i-is}
\end{eqnarray}
In Eq.\ (\ref{eq:i-is}), $\sigma_\nu$ is a spin matrix corresponding to
the direction ${\bf e}_\nu$, which can be any of the Cartesian vectors
${\bf e}_x$,  ${\bf e}_y$, ${\bf e}_z$ or ${\bf e}_{2\theta_\alpha}$,
${\bf e}_{2\theta_\beta}$.

\section{The Numerical Algorithm}\label{sec:dens-op}

The perturbation Hamiltonian $H_{\rm pulse}$, which represents a
laser pulse with two dipolar components of frequencies $\omega_1$ and
$\omega_2$, out of phase with each other by an angle $\phi$, is acting
on the few $n_e = 6$ electrons that occupy the lowest
energy eigenstates. In the discrete representation of the ring, we write
$H_{\rm pulse }(t)=\sum_n H_{\theta}(t)| n \rangle \langle n |$, where
the time dependence $H_{\theta}(t)$ at the point $n$ of the ring is:
\be\label{ht1}
H_{\theta_n}(t)=A e^{-\Gamma t}
[ \sin (\omega_1t)  \cos \theta_n + \sin (\omega_2 t)  \cos (\theta_n+\phi)].
\ee
The pulse of amplitude $A$ is applied at time $t=0$ and lasts for
a time $t_f \sim \Gamma^{-1}$.
The density operator $\rho(t)$ satisfies the differential Liouville equation,
\begin{equation}
i\hbar \dot \rho(t) = \left[ H + H_{\rm pulse}(t), \ \rho(t) \right ] \, ,\label{eq:L}
\end{equation}
whose solutions are obtained numerically, subject to the initial condition\cite{siga}
\be\label{rho0}
    \rho(t=0)=  \sum_{i=1}^{n_e}
         |\Psi_{i}\rangle \langle \Psi_{i}| \, ,
\ee
where only the linear superposition of the lowest $n_e$ occupied states is
considered.  For any $t>0$ Eq.\ (\ref{eq:L}) generates $\rho(t)$ 
through the Crank-Nicholson finite difference
method \cite{gudmundsson,CN} with small time steps $\delta t\ll \Gamma^{-1}$.
The expected value of any observable $O$ of the many-body system is then calculated 
as ${\rm Tr} \left[\rho(t)\hat{O}\right]$, where $\hat{O}$ is the 
single-particle quantum operator associated to that observable.
After the
perturbation ceases to act, at time $t>t_f$, the system remains in an
excited state of constant energy.  The time evolution of the expectation
values of the system observables is determined by their commutation
relation with the Hamiltonian.

In the following calculations the radius of the ring is $r_0=14$
nm, which for $N=20$ sites generates a lattice constant $a\approx 4.4$
nm.  The effective electron mass in the ring is material dependent,
$m^*=0.067m_e$ for a GaAs or $m^*=0.023m_e$ for InAs.  
($m_e$ is the free electron mass.)  Table\,1 summarizes
the various parameters involved in the calculation that pertain to the
applied laser pulse (time, frequencies, pulse amplitude $A$) and those
that pertain to the physical observables of the system (energy, spin,
velocity, charge and spin current $I_c$ and $I_s$).  We also show the
correspondence between the SOI parameters $V_R$ and $V_D$ and Rashba
and Dresselhaus constants $\alpha$ and $\beta$ in the two materials.

\begin{table}[ht]
\centering
{Physical units \\[1ex]}
\begin{tabular}{c c c c}
\hline\hline
\hspace{-2mm}
Parameter &  Unit & GaAs & InAs \\ [0.5ex]
\hline
\hspace{-2mm}
Energy      &\!\!  $V\!=\! {\hbar^2}/({2m^*a^2})$    & 29.4 meV         & 85.6 meV   \\ [1ex]
\hspace{-2mm}
Time               &\!\! $ {\hbar}/{V}$         & 0.022 ps           & 0.0076 ps  \\ [1ex]
\hspace{-2mm}
Frequency          &\!\! $ {V}/{\hbar}$         & 44.6 THz          & 130.0 THz   \\ [1ex]
\hspace{-2mm}
Velocity          &\!\!  $aV/\hbar$             & 196$\cdot10^{12}$ nm/s  & 572$\cdot10^{12}$ nm/s    \\ [1ex]
\hspace{-2mm}
Charge current &\!\!  $eaV/\hbar $    \!                 & 31.5 $ \mu$Anm   & 91.6 $ \mu$Anm  \\ [1ex]
\hspace{-2mm}
Spin current &\!\! ${aV}/{2}$         \!          & 64.6 meVnm    & 188 meVnm \\ [1ex]
\hline
\end{tabular}
\caption{
The specific units for energy, time, frequency, charge
and spin currents, for GaAs ($m^*=0.067m_e$) and for InAs ($m^*=0.023m_e$) quantum rings,
used in the numerical calculations. $e$ is the electron charge.
The discretization constant is a=4.4 nm and is calculated for a ring of
radius $r_0=14$ nm with N=20 discrete points. The charge current is
defined as $I_c=ev$ and spin current is $I_s=sv$, $v$ being the velocity and
$s$ the average spin angular momentum in units of $\hbar$.
}
\label{table1}
\end{table}

In the numerical calculation we use the pulse frequencies
$\omega_1=0.0963V/\hbar$ and $\omega_2=0.276V/\hbar$, while the
attenuation factor and the amplitude are chosen as $\Gamma=4\omega_1$
and $A=2.3V$, respectively.  The two frequencies are close to the first
two Bohr frequencies of the 1D ring calculated in the absence of
SOI, $\omega_{21}=0.0978V/\hbar$ and $\omega_{32}=0.284V/\hbar$. A
Bohr frequency $\omega_{ij}$ is given by the energy difference
$(E_{2i}-E_{2j})/\hbar$, with $E_{2i}$ and $E_{2j}$ being the $2i^{th}$
and $2j^{th}$ eigenvalues of the quantum ring. The factor 2 takes
into account the spin degeneracy. The external pulse last about
$t_f=10/\Gamma$ and is equal to 23$\hbar/V$ dimensionless time units.

According to Table \ \ref{table1}, for a GaAs quantum ring,
the above values correspond to energies $\hbar\omega_1=2.83$ meV,
$\hbar\omega_2=8.11$ meV, the attenuation factor $\Gamma=4\omega_1$,
the amplitude $A=67.7$ meV and the external pulse duration of about
$t_f=0.2$ ps.  If an InAs quantum ring is considered, the numerical
values associated with the laser pulse are $\hbar\omega_1=8.23$ meV,
$\hbar\omega_2=23.6$ meV, $A=197$ meV and $t_f=0.068$ ps.

The strength of the SOI is chosen by using the parameters
$V_\alpha$ and $V_\beta$ in interval $[0,0.08V]$, where $V$ is the energy
unit shown in Table \ref{table1}.  A value $V_\alpha=0.05V$ corresponds to a
Rashba coupling constant $\alpha=12.9$ meVnm in a GaAs quantum ring or
$\alpha=37.6$ meVnm in InAs as shown in Table \ref{table2}.

\begin{table}[ht]
\centering
{Values of Rashba and Dresselhaus parameters \\[1ex]}
\begin{tabular}{c c c}
\hline\hline
Definition  & GaAs & InAs \\ [0.5ex]
\hline
$\alpha=2aV_\alpha$  & $\alpha$=258.4 meVnm$\cdot V_\alpha/V$~~& $\alpha$=752.7 meVnm$\cdot V_\alpha/V$ \\ [1ex]
$\beta=2aV_\beta$    & $\beta$=258.4  meVnm$\cdot V_\beta/V$     & $\beta$ =752.7 meVnm$\cdot V_\beta/V$ \\ [1ex]
\hline
\end{tabular}
\caption{
The Rashba and Dresselhaus parameters $\alpha$ and $\beta$
for GaAs and InAs quantum rings are expressed in terms of the Rashba
and Dresselhaus energies $V_\alpha$ and $V_\beta$, respectively,
in units of $V$. The calculations are done for a ring of radius $r_0=14$
nm with N=20 discrete points.}
\label{table2}
\end{table}

\section{charge and spin currents driven by a symmetric pulse}\label{sec:sym}

\begin{figure}
\includegraphics[width=70mm]{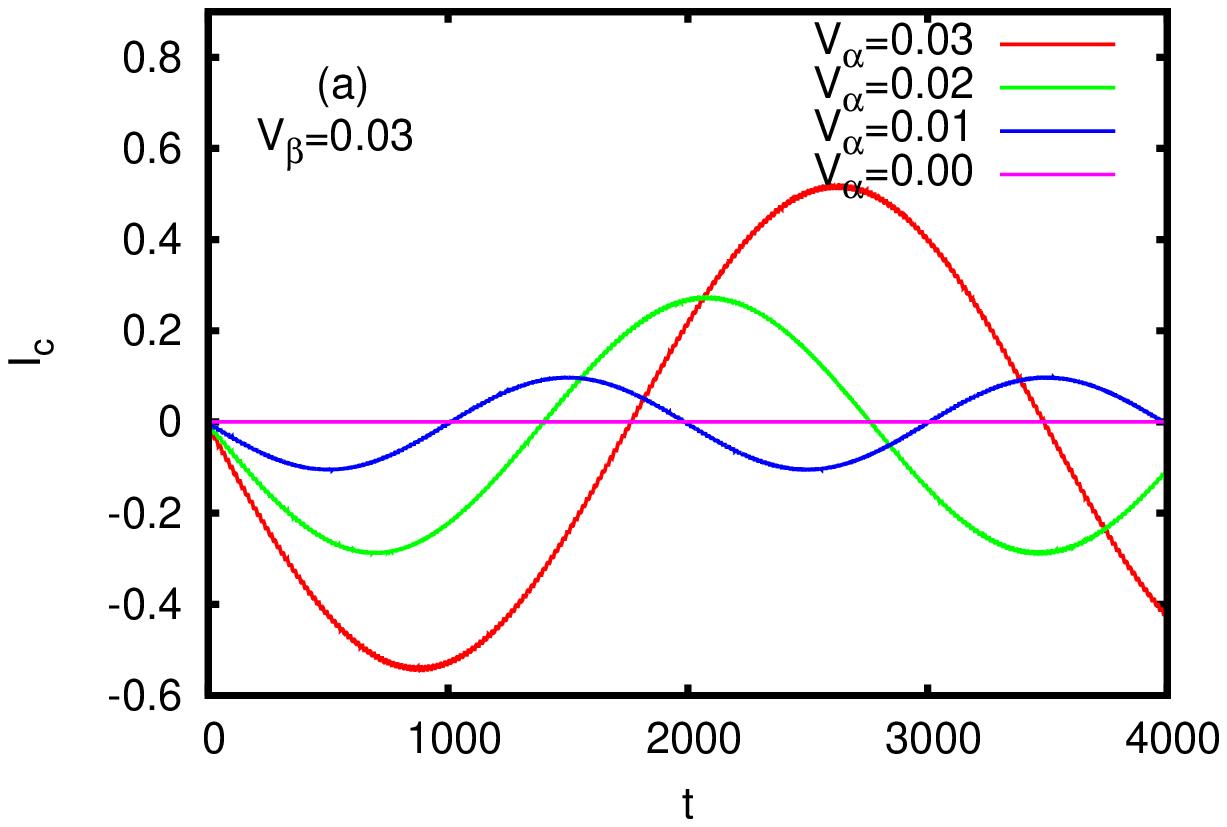}
\includegraphics[width=70mm]{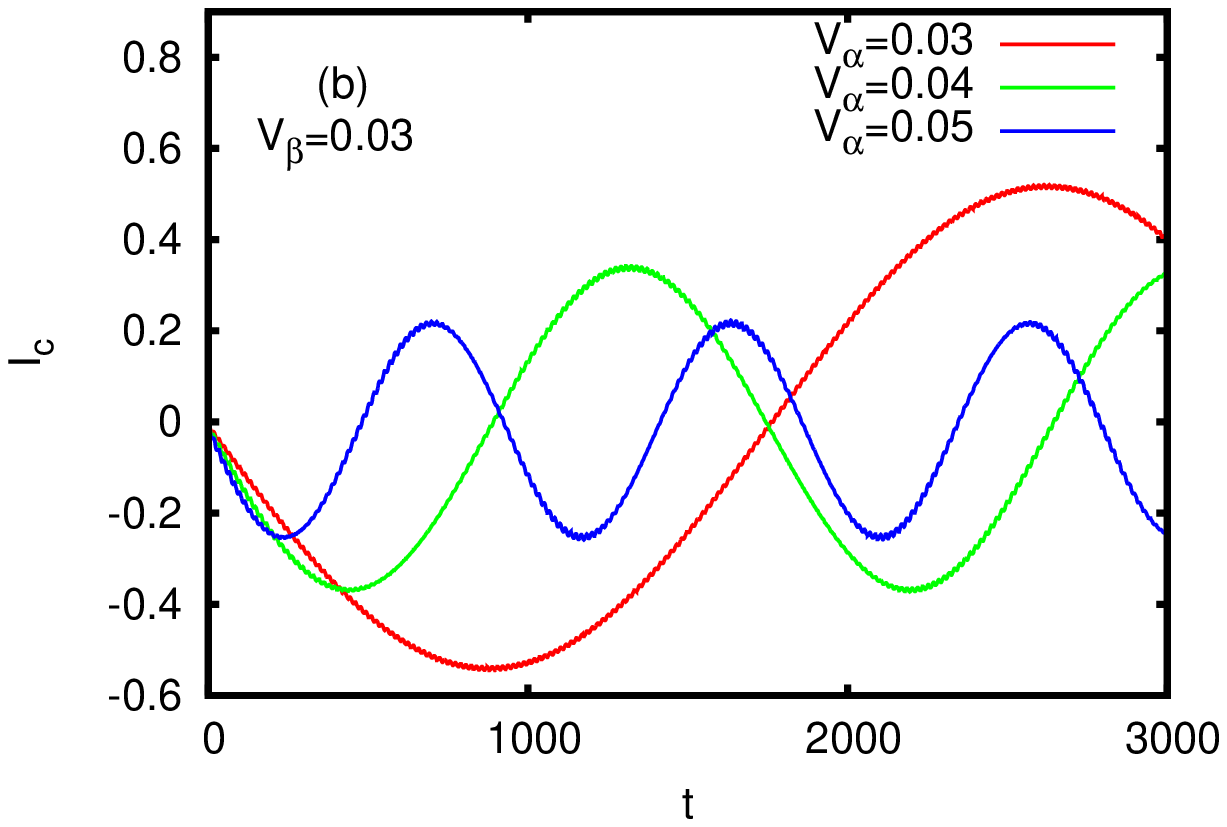}
\includegraphics[width=70mm]{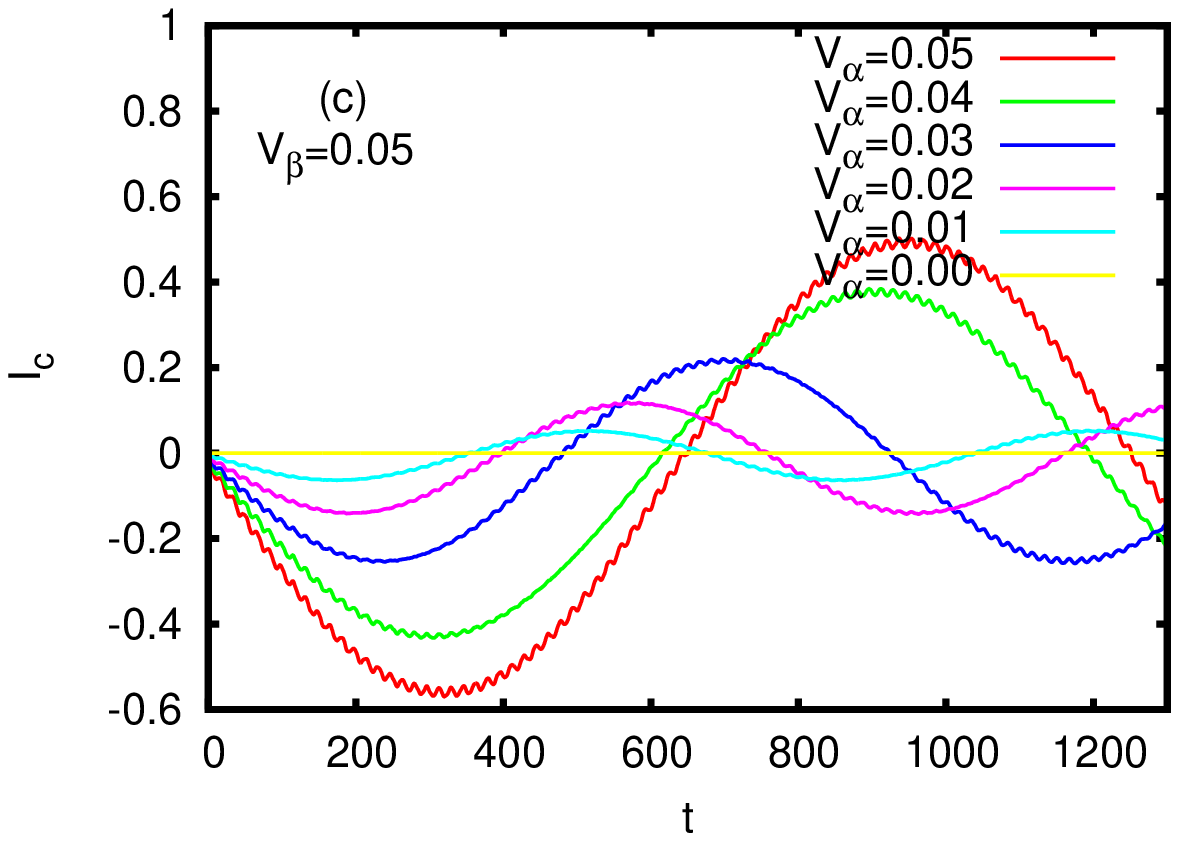}
\includegraphics[width=70mm]{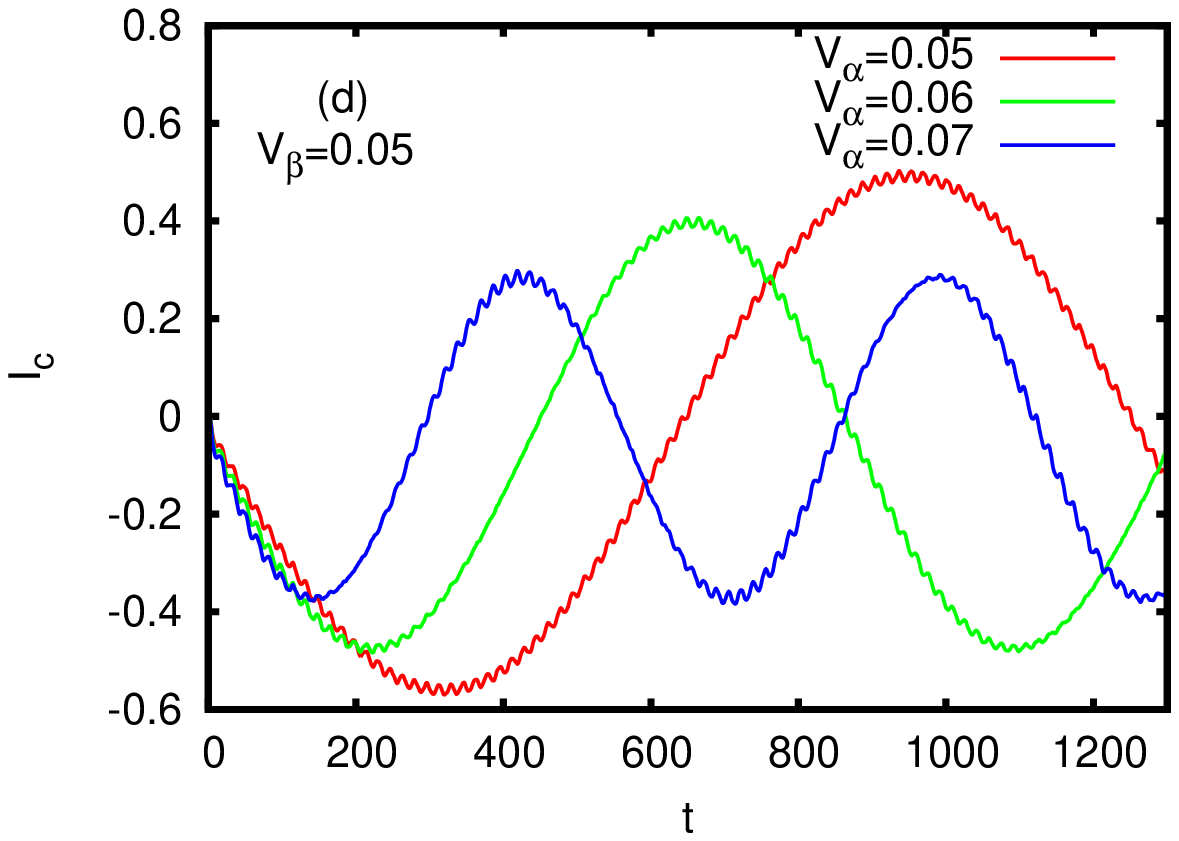}
\caption{(Color online) 
Charge currents $I_c(t)$ in the presence of both Rashba and
Dresselhaus SOI.
$V_{\beta}=0.03$ in (a) and (b) and $V_{\beta}=0.05$ in (c) and (d).
$V_{\alpha}$ is written in the figures.
The minimum frequency $\Omega_{min}=1/T_{max}$ and the maximum amplitude of the oscillations in time, $\Delta I_c$,
are obtained for $V_\alpha=V_\beta$. $\Omega_{min}=0.00028$ for $V_\alpha=V_\beta=0.03V$ and $\Omega_{min}=0.0008$ for $V_\alpha=V_\beta=0.05V$.
The ring and pulse parameters are described in the text. $I_c$ is expressed in $eaV/\hbar$ and time in $\hbar/V$ units.}
\label{charge}.
\end{figure}

We begin our investigation by analyzing the behavior of charge and
spin currents in the presence of an R-D SOI when the ring is
excited by a spatially symmetric laser pulse, i.\ e. $\phi=0$
in Eq.\ (\ref{ht1}). We note that at time $t=0$, the charge current
is zero for all strengths of SOI, $V_\alpha$ or $V_\beta$.  Under the
effect of the perturbation, left-right electron state imbalance occurs,
leading to the establishment of an oscillatory current, as shown in
Fig.\ \ref{charge}, with period $T$ of thousands of time units
$\hbar/V$.  There, the intensity $I_c(t)$ is plotted for various Rashba
couplings when the Dresselhaus coupling assumes two different values,
$V_\beta=0.03V$ in Fig.\ \ref{charge}(a) and Fig.\ \ref{charge}(b),
and  $V_\beta=0.05V$ in Fig.\ \ref{charge}(c) and Fig.\ \ref{charge}(d).
The period and amplitude of the current depend on the SOI parameters,
reaching a maximum when  $V_\alpha=V_\beta$.  The small oscillations of
amplitude $\Delta' I_c < 0.012\ eaV/\hbar$ and period $T'\approx 16\ \hbar/V$
noticeable in Figs.\,\ref{charge}(c) and \ref{charge}(d) are further
investigated in the discussion of the spin currents.

\begin{figure}
\includegraphics[width=70mm]{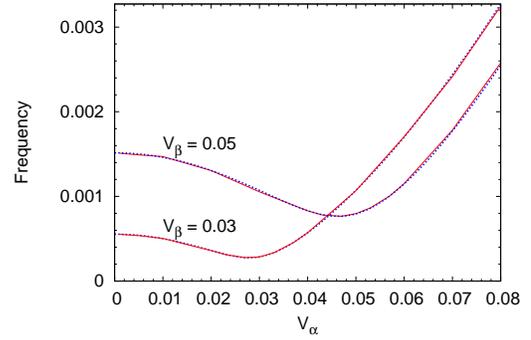}
\caption{(Color online) 
The frequency of the long oscillations of the charge current
are plotted as functions of the Rashba energy $V_\alpha$ for two values of
the Dresselhaus coupling $V_{\beta}=0.03V$ and $V_{\beta}=0.05V$ respectively.
The continuous lines are obtained from the time dependent output data shown in
Fig.\ \ref{charge}.  The dotted lines show the results for the Bohr frequency
$\omega_{32}/2\pi\equiv (E_6-E_4)/(2\pi)$ (see text).
The minimum frequency is obtained for $V_\alpha=V_\beta$.
The frequency unit is $V/\hbar$ and $V_\alpha$ is in units of V.
}
\label{freq32}
\end{figure}

The low frequency of the charge current $\Omega=1/T$ is given
by the Bohr frequency corresponding to the occupied states with
highest energies in the unperturbed ring, $E_6$ and $E_4$, such
that $\Omega=\omega_{32}/2\pi$.  We verified this result for all the
SOI strengths used in these calculations.
For example, for $V_{\alpha}=0.05$ and $V_{\beta}=0.03$ the first few
(relevant) energy levels of the quantum ring are, in the increasing order,
$E_1=E_2=-0.003383,\ E_3=E_4=0.09131,\ E_5=E_6=0.09805,\ E_7=E_8=0.3735,\
E_9=E_{10}=0.3850,\ E_{11}=E_{12}=0.8146$.  We thus obtain from the
energy spectrum $T=2\pi/\omega_{32}\equiv 2\pi/(E_6-E_4)=932$ time units,
whereas directly from the numerical results of the time dependent current we
get $T=948$.  This means that after the
perturbation had ceased, the ring remains in an excited state which is a
superposition of states with partial population of the energy levels
$E_6$ and $E_4$.  The frequency of the oscillations, vs. the Rashba SOI
$V_\alpha\in[0,0.08V]$, is plotted in Fig.\ \ref{freq32}, for two values of
the Dresselhaus SOI.  Both the results obtained directly from the time dependent
charge current, and from the Bohr frequency $\omega_{32}$, are shown.  The
agreement is almost perfect.  We find a non-monotonic behavior that is highly
dependent on the relative strengths of the two SOI's, such that for
a fixed $V_{\beta}$, the minimum frequency occurs for
equal strengths $V_{\alpha}=V_{\beta}$.  In this case additional
degeneracies exist in the energy spectrum,\cite{egues} and therefore
when $V_{\alpha}\to V_{\beta}$ all energy gaps tend to shrink.  (With our
parameters, for equal SOI strengths, we obtain $E_{15}=E_{16}=E_{17}=E_{18}$.)

The fast oscillations of the charge current, which can be seen well
in Figs.\ \ref{charge}(c-d) can also be explained by the energy spectrum.
They are created by transitions between states separated by relatively
large energies, like $E_6$, $E_4$, $E_2$ and $E_8$ or $E_{10}$. In fact
the period of the fast oscillations corresponds to the largest energy
gap in this series, $T'=2\pi/(E_{10}-E_2)\approx 16\ \hbar/V$.


\begin{figure}
\includegraphics[width=70mm]{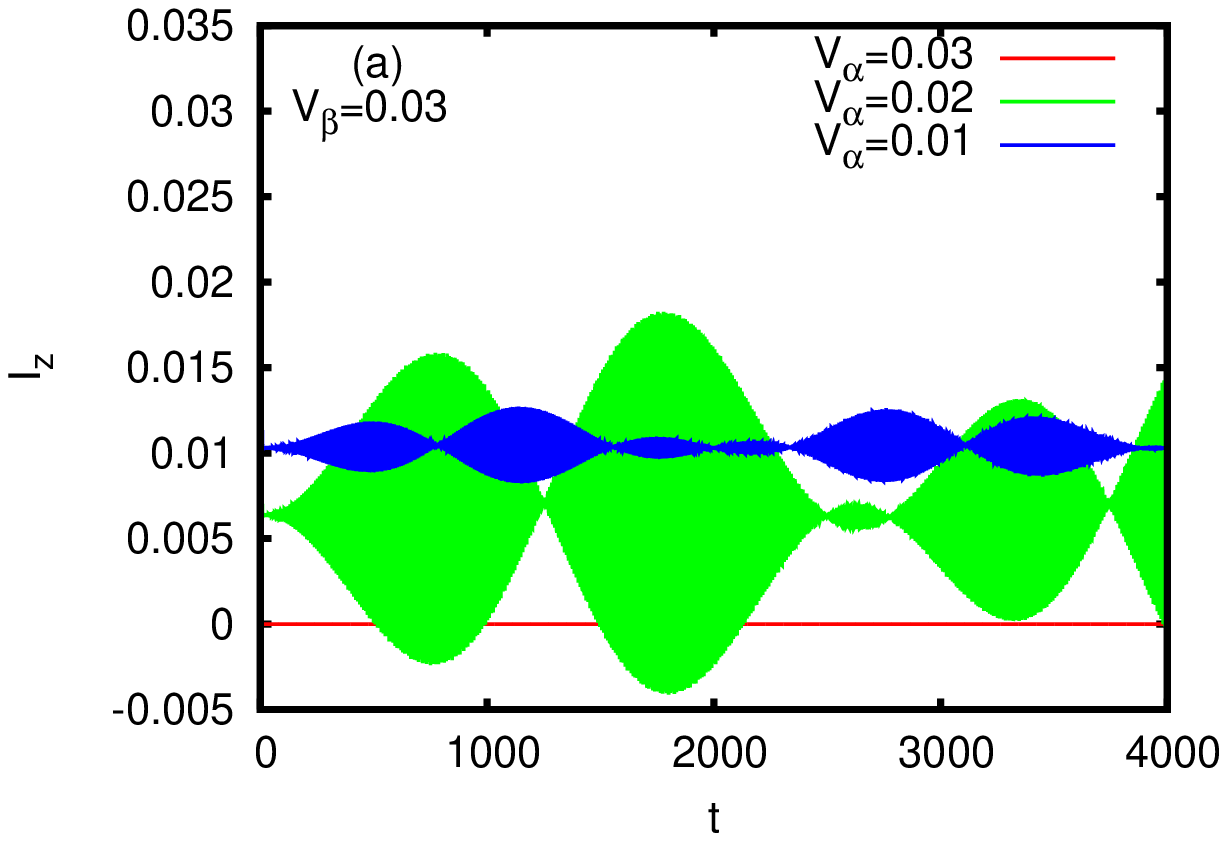}
\includegraphics[width=70mm]{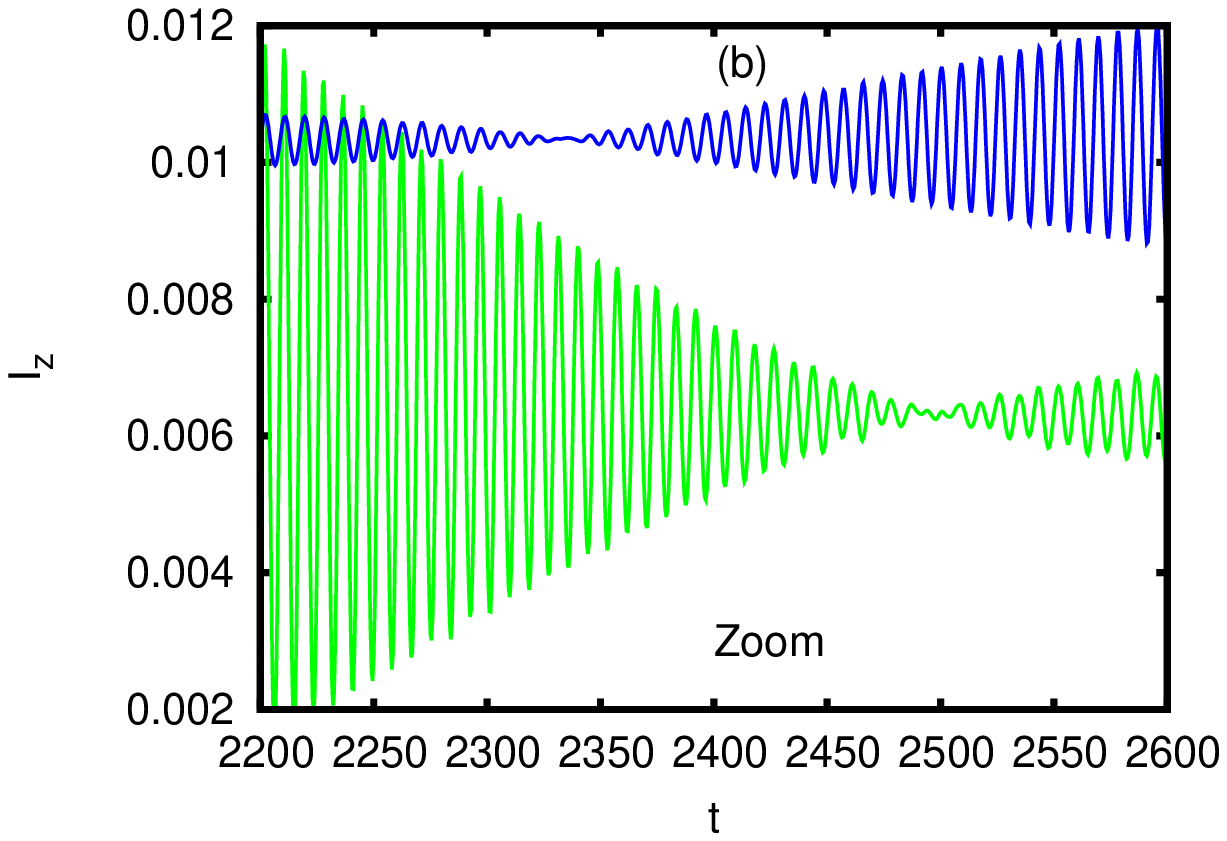}
\includegraphics[width=70mm]{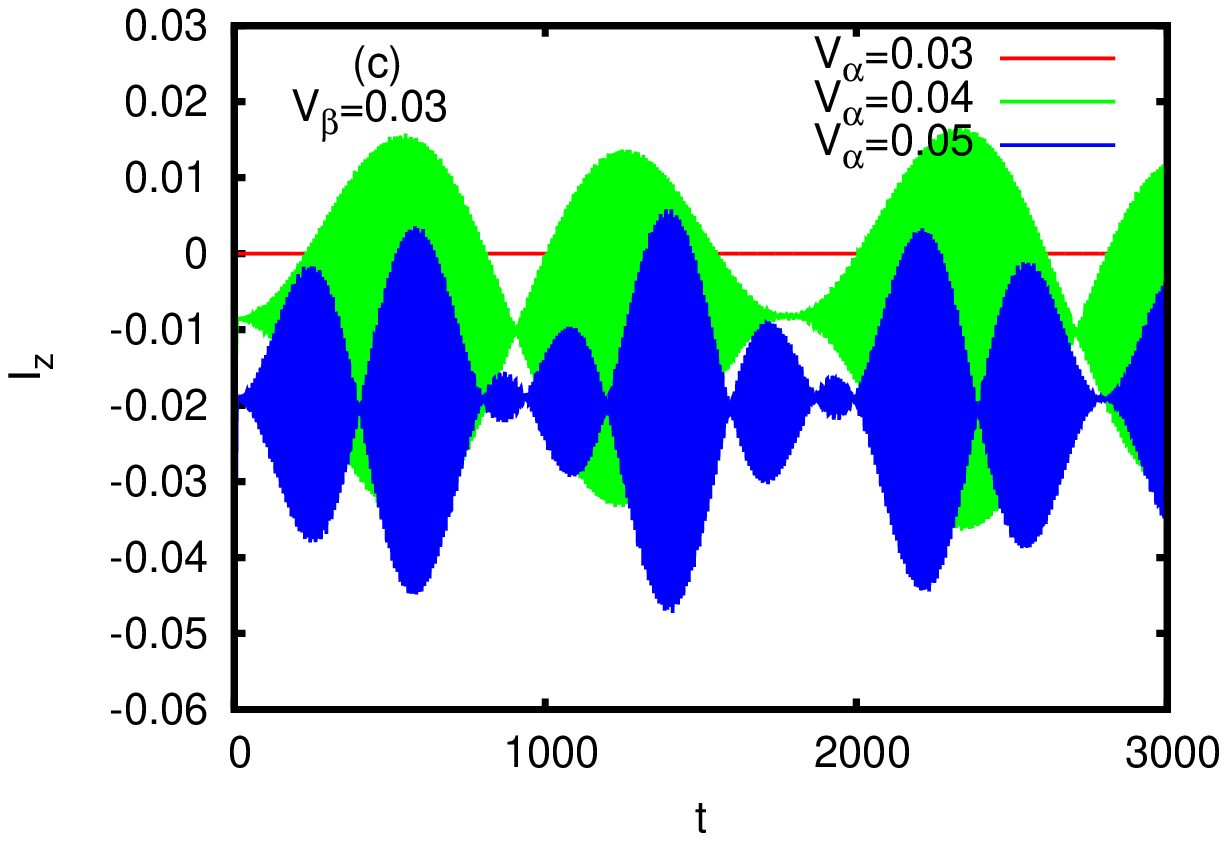}
\includegraphics[width=70mm]{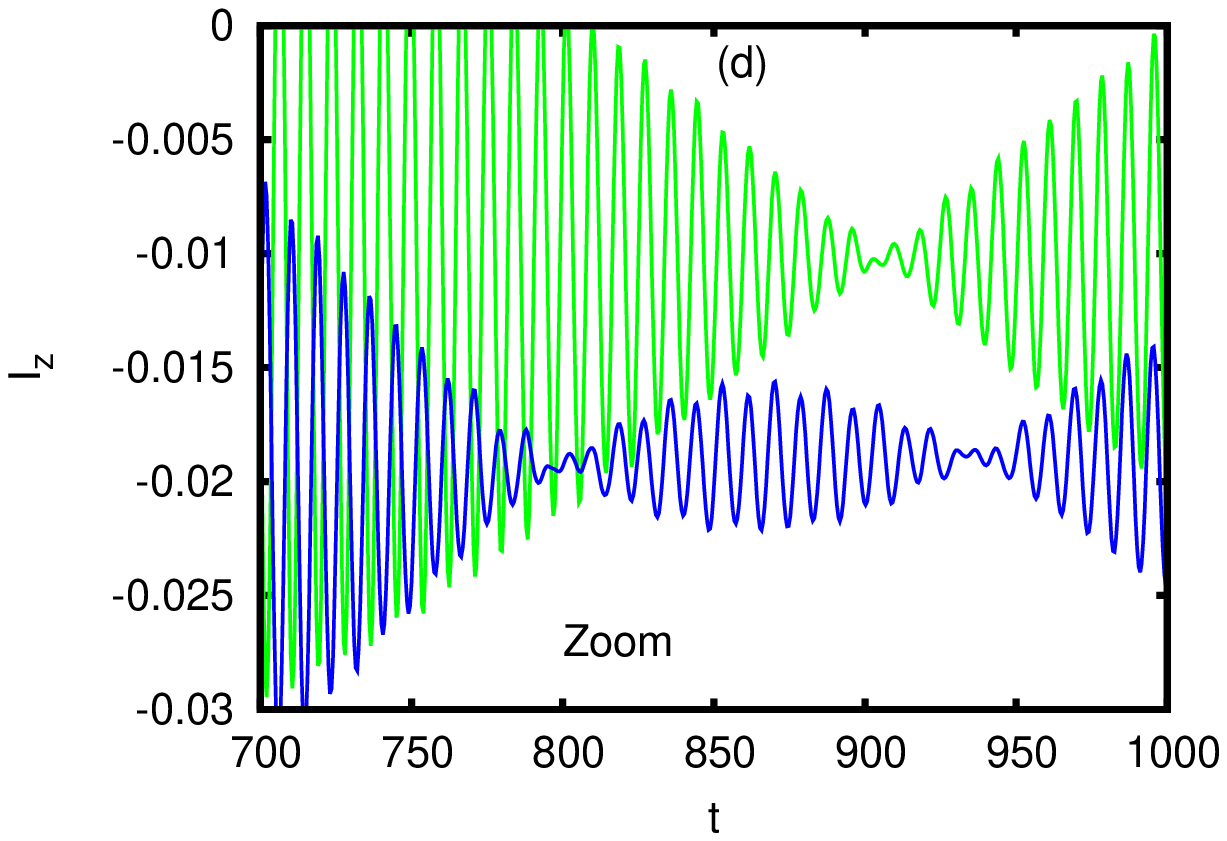}
\caption{(Color online) 
The time variation of the spin current $I_z(t)$  in the presence of
Rashba and Dresselhaus SOI. The Dresselhaus energy is $V_{\beta}=0.03$ in (a)
and (b). The Rashba energy is $V_{\alpha}=0.01, 0.02 ,0.03$ in (a) and
$V_{\alpha}=0.03, 0.04,0.05$ in (b).  For $V_\alpha=V_\beta$ the spin current
is zero. For $V_\alpha\ne V_\beta$ there are spin current oscillations.
Figures (c) and (d) present a magnified view of (a) and (b) respectively.
(Spin current unit is $aV/2$ and time unit is $\hbar/V$.)}
\label{iz56}
\end{figure}

Under the same circumstances we study the time variation of the
spin current associated to the spin projection along
the $\hat{z}$-direction, $I_z(t)$, for various values of the
SOI interactions. In this instance, the oscillatory behavior
associated with charge displacement is complicated by a a beating
pattern with nodal points of zero oscillation, as seen in Figs.\
\ref{iz56}(a) and (c). Zooming in near the nodal points, we obtain
Fig.\ \ref{iz56}(b) and \ref{iz56}(d), where it is readily observable
that, as it passes through a nodal point, the spin current phase changes by
$\pi$. As before, the amplitude of the oscillations depends on the two
SOI interactions and their ratio, while the period of the fast
oscillations $T'$ seems quite insensitive to this aspect.  The period of
the fast oscillations is $T'\approx 8\ \hbar/V$ or slightly more
for all examples shown in Figs.\ \ref{iz56}(a-b), while the amplitude
varies from zero, at nodal points, to about $0.1\ aV/2$. The period
$T'$ of the spin current is actually half of the period of the fast
oscillations of the charge current (also denoted by $T'$ in the analysis
of that current, and found there $\approx 16$ time units).  The reason is that
the spin precesses relatively to two principal axes, independently.
The axes are tilted at different angles $2\theta_{\alpha}$ and
$2\theta_{\beta}$, due to the Rashba and Dresselhaus SOI, respectively.
The frequency of the spin current is thus roughly the double of that
for the charge current, because of the two spin modes.\cite{nita2} For
$V_{\alpha}=V_{\beta}$ the spin oscillations cancel each other and the
spin current vanishes, as seen in Figs.\ \ref{iz56}(a) and (c). For
$V_{\alpha}\neq V_{\beta}$ the spin current actually has an {\em aperiodic}
time evolution, observable in the same mentioned figures. So we cannot
really speak about a rigorous periodicity of the spin current, and
consequently of the charge currents as well.

We interpret the beating pattern of the spin current oscillations by
a nutation motion of the electron spin between the two quantization
directions ${\bf e}_{2\theta_\alpha}$ and ${\bf e}_{2\theta_\beta}$
imposed by the SOI couplings that force the electron spin precession
in opposite directions.  When only one type of the SOI, say Rashba,
is present, ${\bf e}_{2\theta_\alpha}$ becomes a good quantization axis
for spin.  The corresponding spin current $I_{2\theta_\alpha}$ commutes
with the equilibrium Hamiltonian and consequently, remains constant in
time as a conserved observable.  When the  SOI parameters are not equal,
the principal spin axes have different angles $2\theta_{\alpha}\neq
2\theta_{\beta}$ and the electron spin executes rapid oscillations
between them.

\begin{figure}
\includegraphics[width=3in]{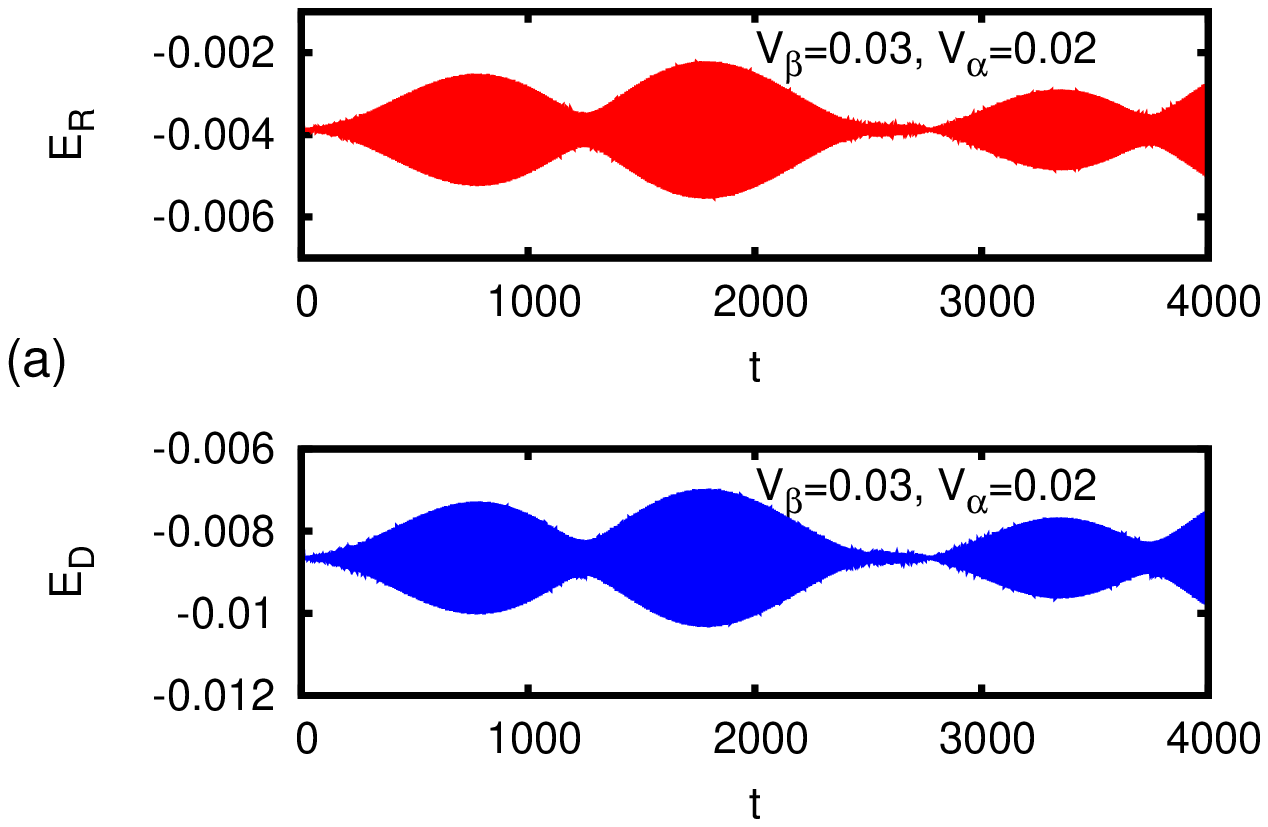}
\includegraphics[width=3in]{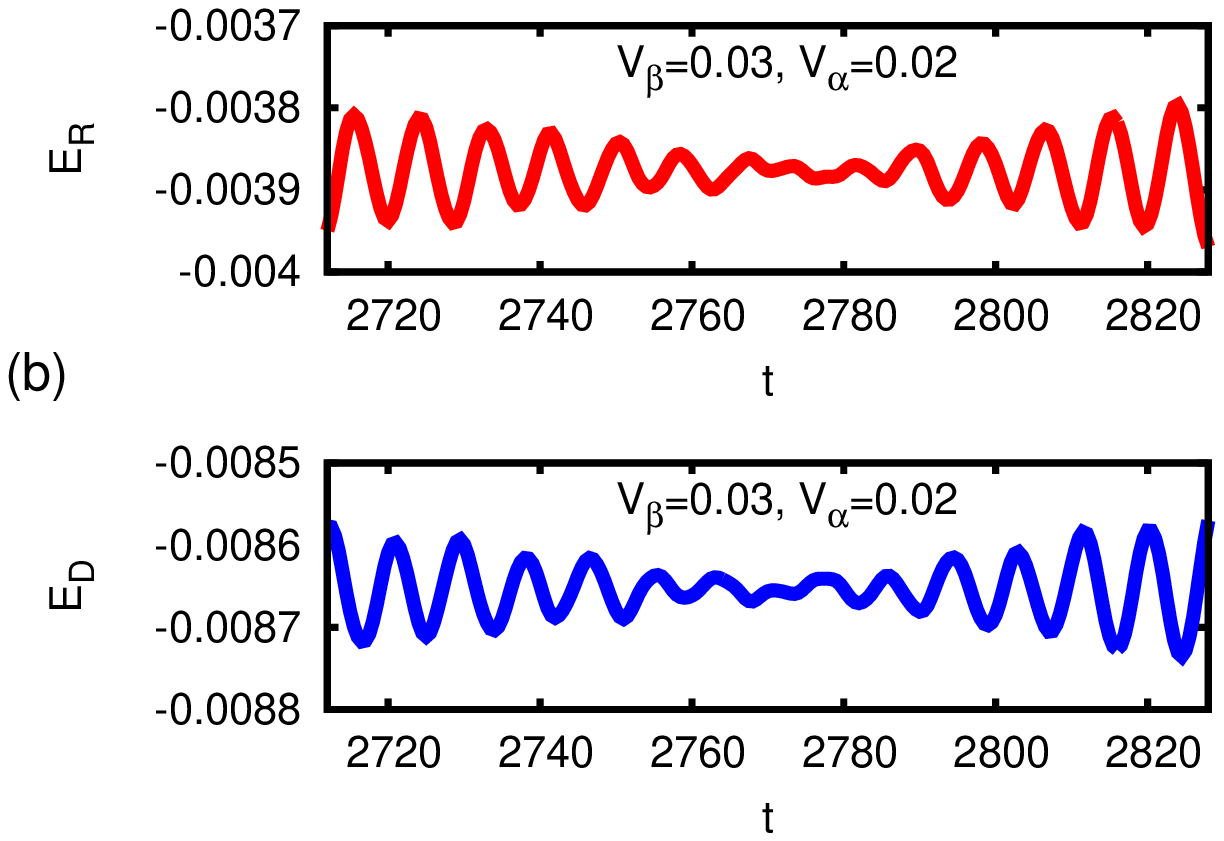}
\caption{(Color online) 
The out of phase oscillations of the Rashba and Dresselhaus
potential energies, $E_R(t)$ and $E_D(t)$.  $V_{\beta}=0.03V$
and $V_{\alpha}=0.02V$,  (a) for the complete time interval, and
(b) zoomed into a small time interval. The energy unit is V and time unit
is $\hbar/V$.  }
\label{energy1}
\includegraphics[width=3in]{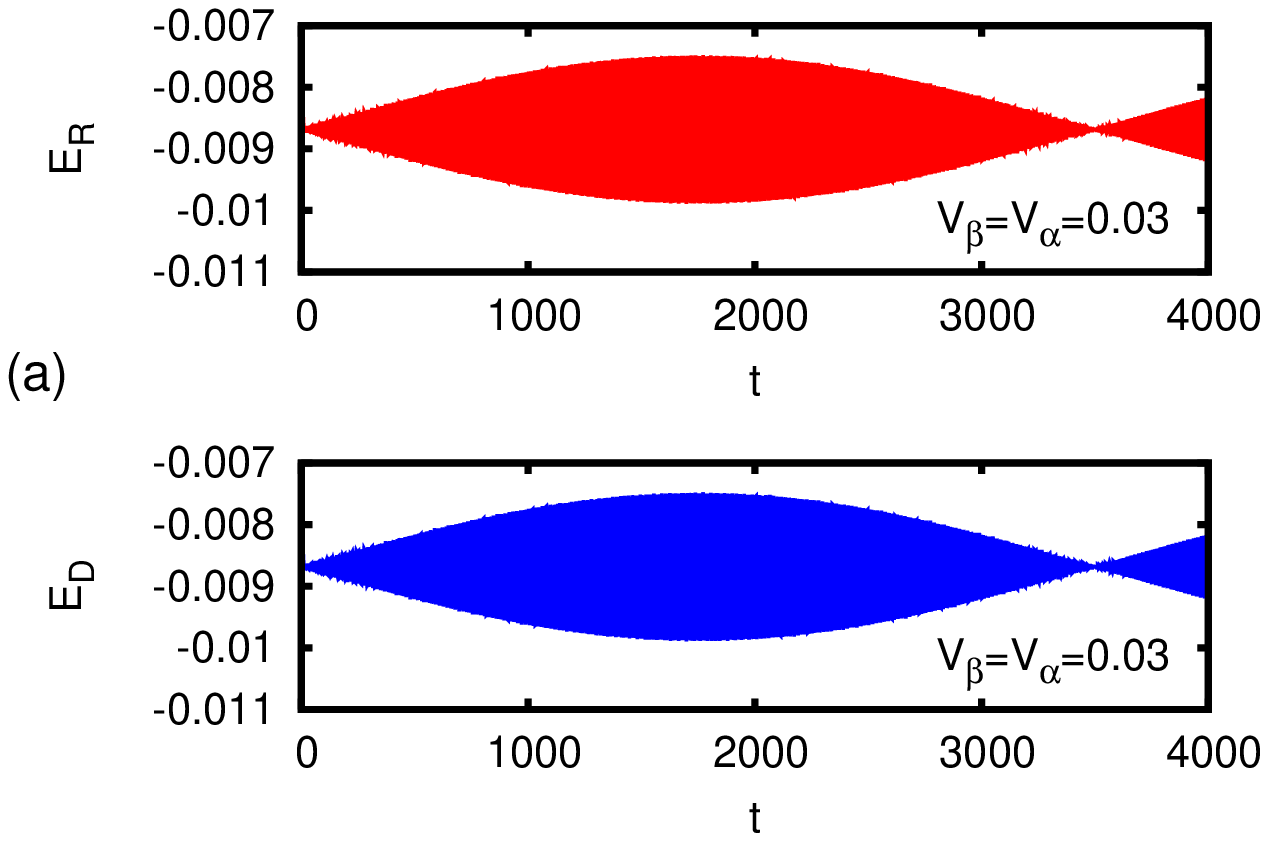}
\includegraphics[width=3in]{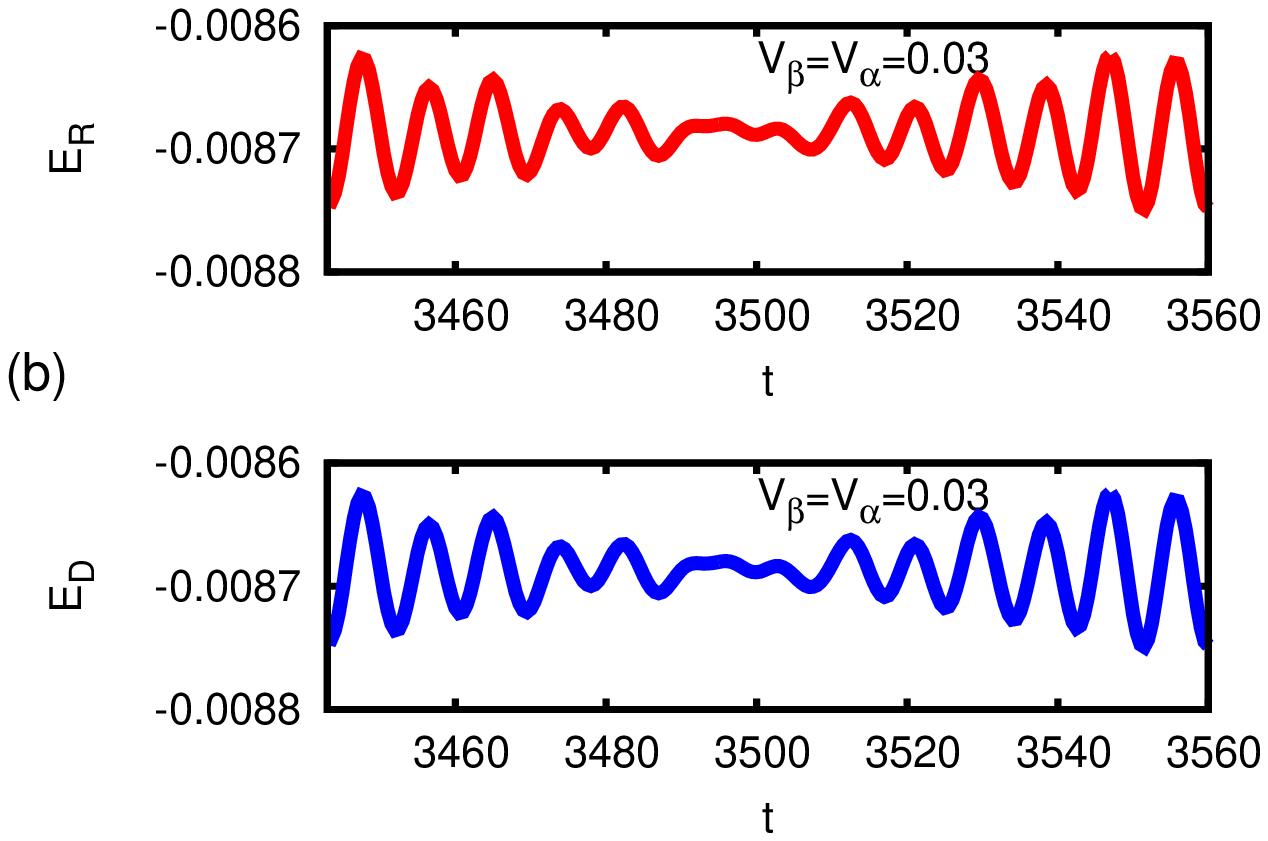}
\caption{(Color online) 
The in phase oscillations of the Rashba and Dresselhaus potential
energies, $E_R(t)$ and $E_D(t)$, for
$V_{\beta}=V_{\alpha}=0.03V$,  (a) for the complete time interval, and
(b) zoomed into a small time interval. The energy unit is V and time unit
is $\hbar/V$.  }
\label{energy2}
\end{figure}

Another perspective on this phenomenon is revealed by plots of the time
oscillation of the SOI energy, $E_R$ and $E_D$, defined as the expected
values of the potential energies $V_R$ and $V_D$ given by
Eqs.\ (\ref{vr}-\ref{vd}).  Figs.\ \ref{energy1}(a) and \ref{energy1}(b)
display $E_R(t)$ and $E_D(t)$ for different values of SOI, while in
\ref{energy2}(a) and \ref{energy2}(b) the same value $V_\alpha=V_\beta$
is used.  It appears that an exchange between Rashba and Dresselhaus
potential energies occurs with the same oscillation period as the $I_z$
beat oscillations.  The orbital motion of the electron is thus accompanied
by vibrations of the spin orientation between these two directions that
give rise to out of phase oscillations of Rashba and Dresselhaus energy.
At equal SOI parameters, $V_{\alpha}=V_{\beta}$, the two potential
energies reach the same value and oscillate in-phase. Since the precession
of the electron spin around the two preferential directions or the Rashba
and Dresselhaus SOI's occurs with equal amplitudes in opposite directions,
the spin current vanishes.

\section{Charge and Spin currents driven by an asymmetric pulse}\label{sec:asym}

The time evolution of the electronic states in the quantum ring 
excited by an asymmetric pulse is asymmetric is discussed next. This involves
the full form of Eq.\ (\ref{ht1}), with $\phi\neq 0$.  In this
case, we are interested in establishing the geometric effect that the
dephasing angle $\phi$, varied in fixed steps $\phi=0$, $\pi/6$, $2\pi/6$,
$\cdots$, $11\pi/6$, has on the charge and spin currents. To better
focus on this aspect, the Rashba and Dresselhaus interactions 
are fixed at values $V_\alpha=0.02V$ and $V_\beta=0.03V$, respectively. In our
analysis, two dynamic regimes are considered, delimited by the lifetime of
the perturbation, $t_f$, one occurring for $t<t_f$, and the other one for $t>t_f$
when the persistent oscillatory behavior is established.

\begin{figure}
\includegraphics[width=57mm]{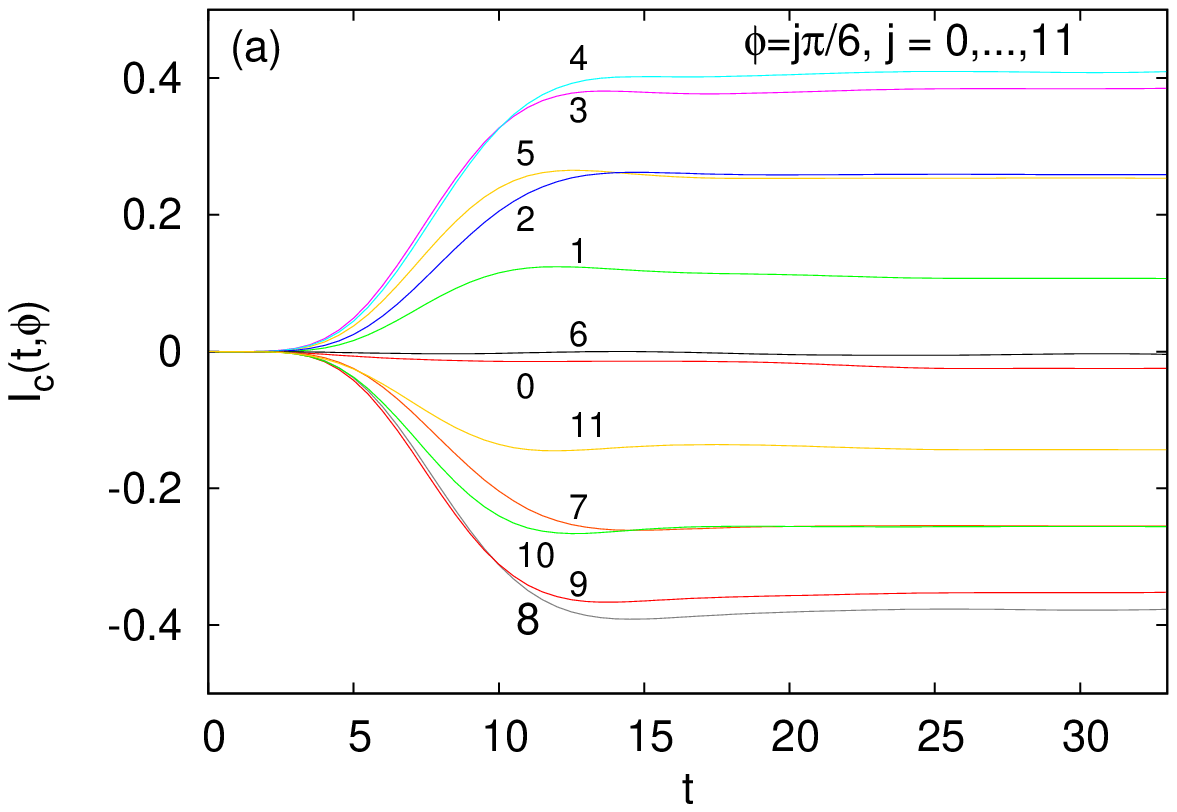}
\includegraphics[width=57mm]{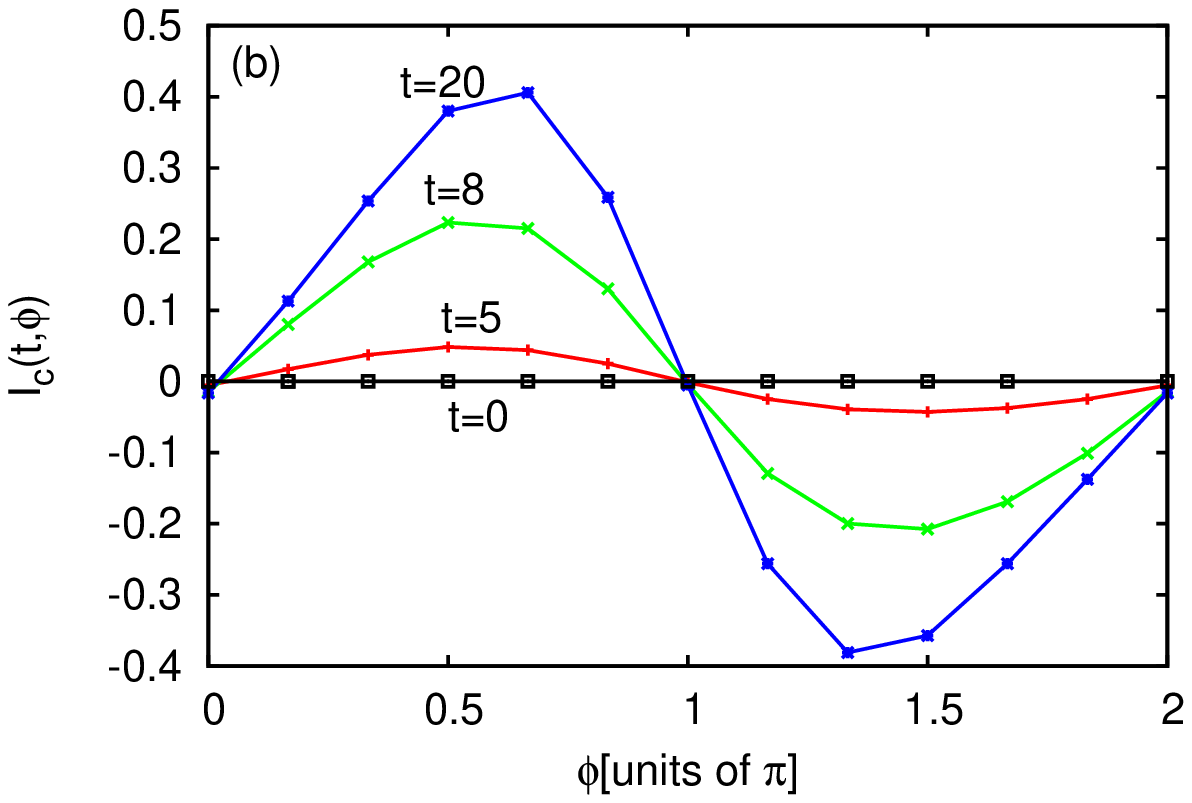}
\includegraphics[width=57mm]{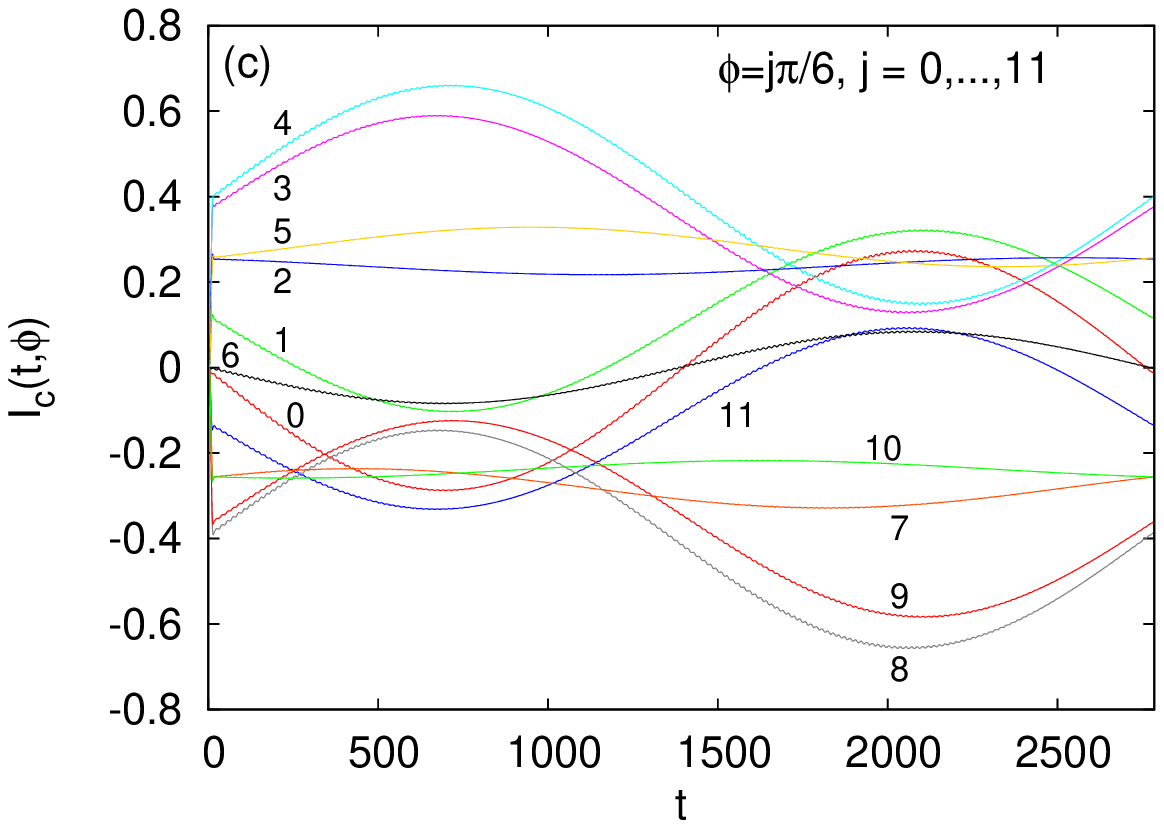}
\includegraphics[width=57mm]{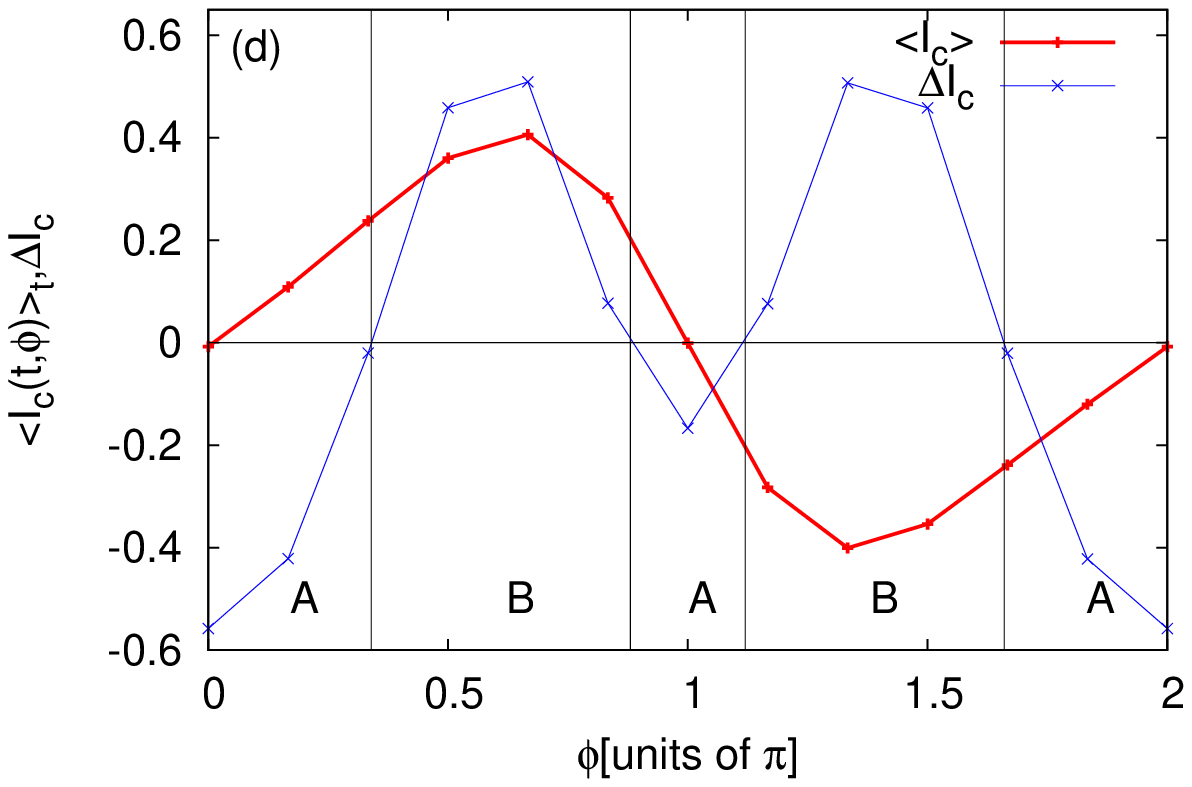}
\vspace{-5mm}
\caption{(Color online)  (a) The charge current  $I_c(t,\phi)$ as function of
time during the radiation pulse for $t\in[0,33]$, for different
angles $\phi=j\pi/6$ between the two dipoles. The integer $j=1,2,...,11$
is indicated near each curve.  In (b) we show the charge current as a
function of $\phi$ at several moments shown in the graph, $t=0,5,8$ and $20$.
(c) The time evolution of the current over a full period of the oscillation
$T\approx 2783$.  The angles $\phi$ are the same as in (a). (d) The
average charge current $\langle I_c \rangle_t$ and the amplitude $\Delta I_c$
are shown versus the angle $\phi$. The time average is
$\langle I_c \rangle_t=[I_c(T/4)+I_c(3T/4)]/2$ and the amplitude is $\Delta
I_c=I_c(T/4)-I_c(3T/4)$. The intervals of type A and type B oscillations are
indicated.  The SOI parameters are $V_\alpha=0.02V$ and
$V_\beta=0.03V$.  Time physical units are $\hbar/V$ for the time and
$eaV/\hbar$ for the current.
}
\label{icas}
\end{figure}

In Fig.\ \ref{icas}(a) the charge current $I_c(t,\phi)$ is shown
to evolve under the effect of the two-component laser pulse
between $0<t<t_f$, from zero at $t=0$ to non-zero values at
$t=t_f$, for several angles $\phi$.  The largest currents occur for
dephasing angles $\phi=4\pi/6$ and $8\pi/6$, whereas the minima occur for
$\phi=0$ and $\pi$. A similar pattern is seen in Fig.\,\ref{icas}(b),
where the dependence of the instantaneous charge current at a given time
$I_c(t,\phi)$ is depicted for various angles $\phi$.  A maximum variation
is noticeable at $\phi=\pi/2$ and $3\pi/2$ only for $t=5\hbar/V$ and
$t=8\hbar/V$, while for larger times, such as at $t=20\hbar/V$
for instance, the maximum variation of the induced current $|I_c(t)|$
is realized for dephasing angles $\phi=4\pi/6$ and $8\pi/6$ as already seen
in Fig.\ \ref{icas}(a).

After the radiation pulse vanishes, i.\ e. for $t>t_f\simeq
25\hbar/V$, the charge current oscillates with a period $T\simeq
2783 \hbar/V$, as shown in Fig.\ \ref{icas}(c), where now the
time covers one complete period.  The amplitude of the oscillations
is strongly affected by the value of the dephasing angle. For the
dephasing angles $\phi=0$, $\pi/6$ and $11\pi/6$ the time dependence
of the charge current has the shape of the function
$-\sin(2\pi t/T)$, and will be called here type A oscillations.
For the angles $\phi=3\pi/6$, $4\pi/6$, $8\pi/6$, and $9\pi/6$ the
oscillations look like $\sin(2\pi t/T)$ and will be
called type B oscillations. Type B are shifted from type
A by half a period, $T/2$. Type A oscillations of much smaller
amplitudes are also seen in Fig.\ \ref{icas}(c) for $\phi=6\pi/6$,
and also for $\phi=2\pi/6$ and $10\pi/6$.  The latter two
are actually asymmetric and in transition from type A to type B.
For $\phi=5\pi/6$ and $7\pi/6$ again small amplitudes are
obtained, this time for oscillations of type B.  The amplitude of
the current, defined as $\Delta I_c=|I_c(T/4)-I_c(3T/4)|$, changes with
$\phi$ as shown in Fig.\ \ref{icas}(d). As the angle $\phi$ is varied,
the two types of oscillations alternate. This result defines four critical angles
$\phi=\phi_{c1,c2,c3,c4}$ that correspond to a crossover from type A to
type B.  Their numerical values obtained by the interpolation
of the presented curves are: $0.35\pi$, $0.85\pi$, $1.15\pi$, and
$1.65\pi$, respectively.  At the A to B crossover, \i.\ e. at
the critical angles $\phi_{c1,\dots,c4}$, the oscillations of the charge
current vanish, as shown in Fig.\ \ref{icas}(d). This means that,
by tuning the value of the dephasing angle of the external pulse
such that $\phi=\phi_{c1,\dots,c4}$ the ring can be excited into a
final state that supports a non-zero charge current with zero amplitude,
although the current operator $I_c$ is a nonconservative observable in
the presence of both Rashba and Dresselhaus interactions.  The time
average of the charge oscillations, defined as
$\langle I_c \rangle_t=(I_c(T/4)+I_c(3T/4))/2$, is also shown
in Fig.\ \ref{icas}(d) as function of $\phi$. We note that, for type
A oscillations, $\langle I_c \rangle_t=0$ for angle $\phi=\phi_m=0$
and $\pi$, and acquires its maximum and minimum values of $\pm0.4$ at
the angle $\phi=\phi_M=0.6\pi$ and $\phi=\phi_{-M}=1.4\pi$, for B type
oscillations.
This suggests that, by tuning the dephasing angle
$\phi$, both the type of charge current oscillations, as well as their
time average can be modified.


\begin{figure}
\includegraphics[width=57mm]{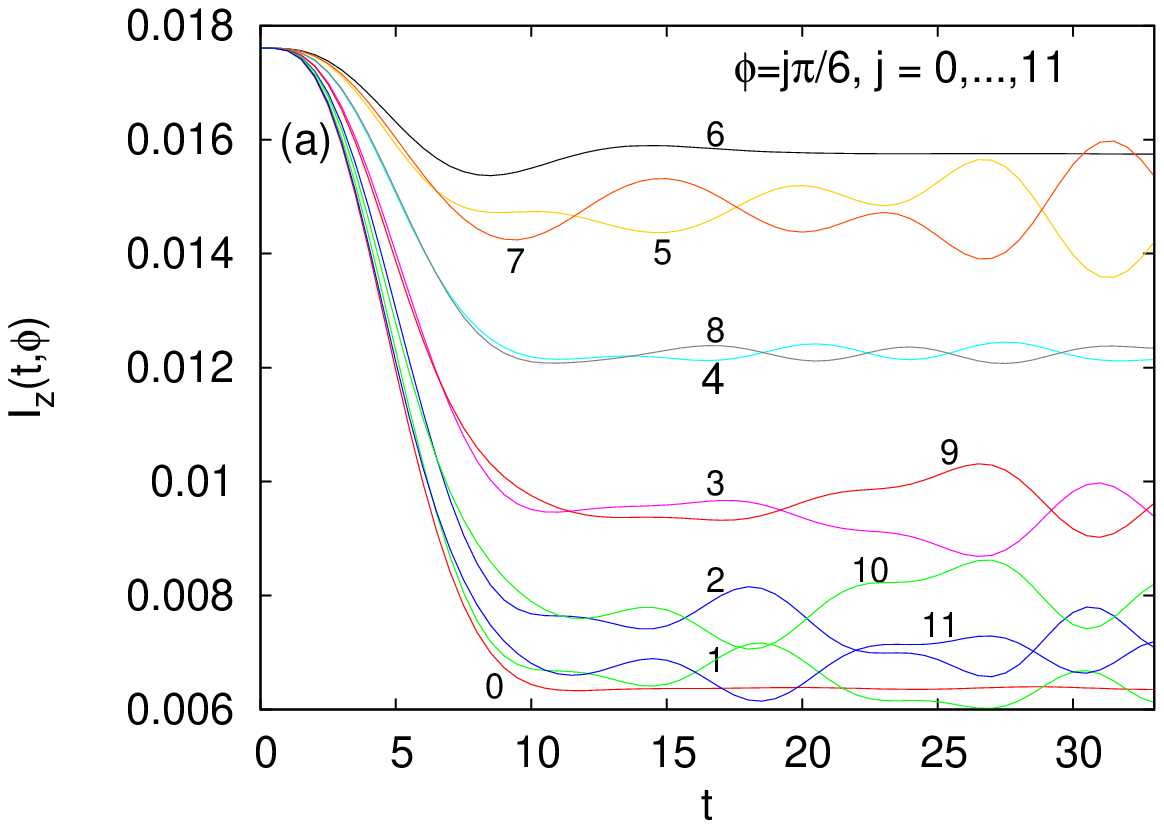}
\includegraphics[width=57mm]{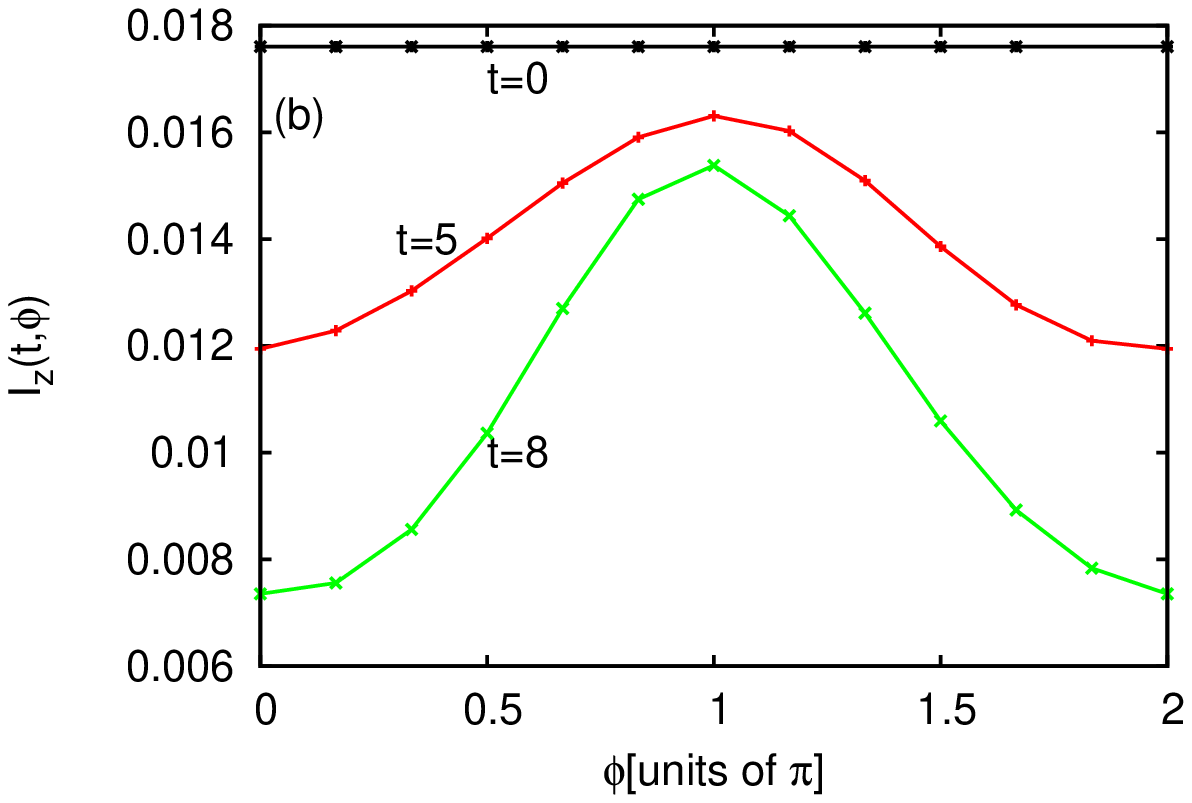}
\includegraphics[width=57mm]{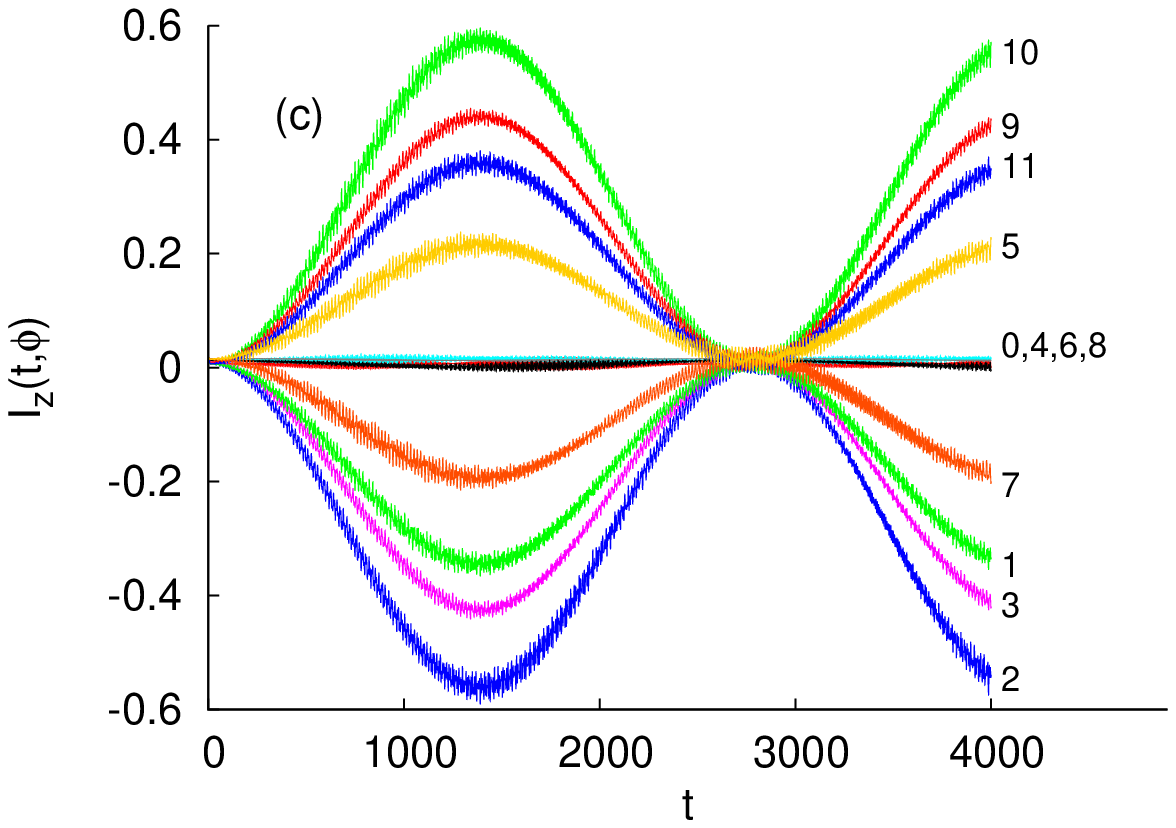}
\includegraphics[width=57mm]{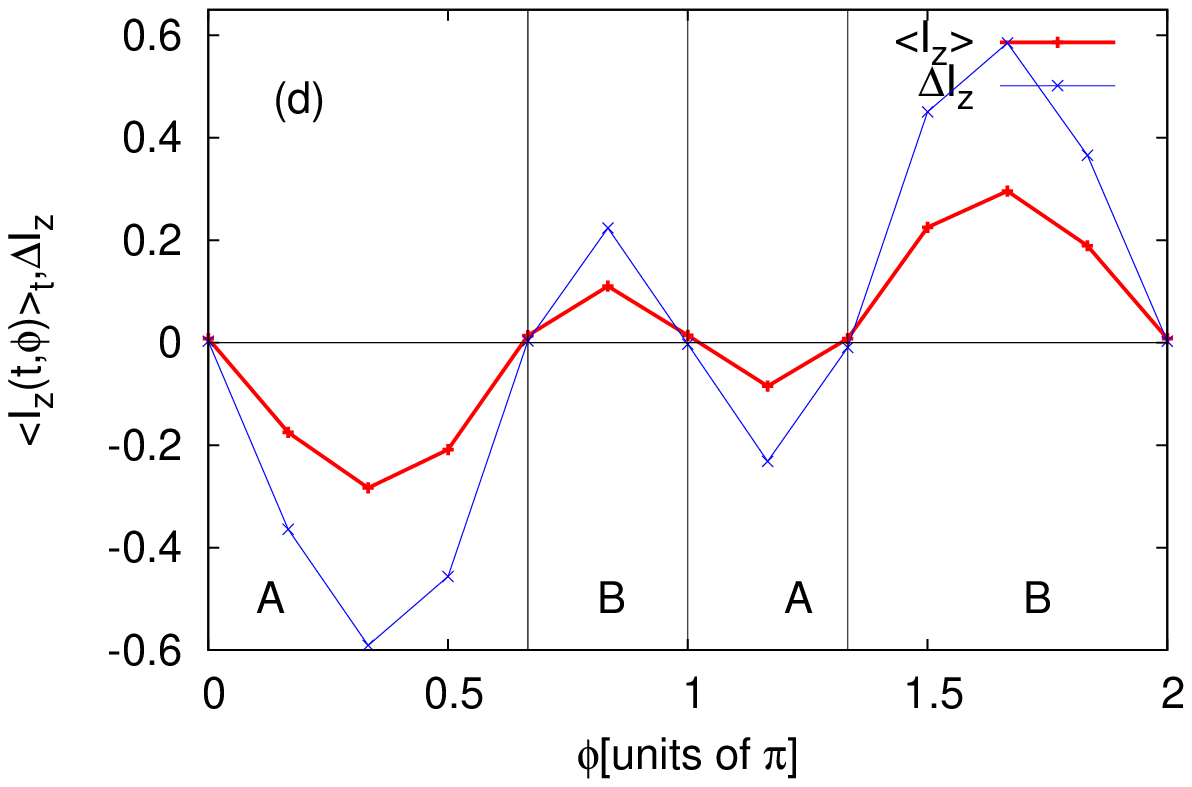}
\vspace{-5mm}
\caption{(Color online) 
(a) The spin current $I_z(t,\phi)$ during the pulse, for
$t\in[0,33]$ in (a) for the same series of angles between the dipoles as
in Fig.\ \ref{icas}(a).  In (b) we plot the spin current vs. the angle
$\phi$ at several moments, t=0, 5 and 8. (c) The spin current along the
$z$ direction within a complete period $T$.  The spin currents have long
oscillations with period $T\simeq 2783$ and short oscillations with
$T'\simeq 8$.  The integers $j=1,2,...,11$ near the curve correspond to
the angles $\phi=j\pi/6$.  In (d) we summarize the results by showing
the time average of the spin current $\langle I_z\rangle_t=(I_z(T/2)+I_z(T))/2$
and its amplitude $\Delta I_z=I_z(T/2)-I_z(T)$ versus the dephasing
angle $\phi$ and we indicate oscillations of type A and type B. The SOI
parameters are $V_\alpha=0.02V$ and $V_\beta=0.03V$.
Time unit is $\hbar/V$ and the unit for the spin current is $Va/2$.
}
\label{isas}
\end{figure}

In a parallel analysis, we present the time evolution of the spin current
in Fig.\,\ref{isas}. At time $t=0$, the ring is found in the ground
state, which has a permanent spin current $I_z(t=0)=0.01760Va/2$. Along the SOI
principal axes the spin currents are $I_{2\theta_\alpha}(t=0)=0.01775Va/2$
and $I_{2\theta_\beta}(t=0)=0.01793Va/2$, respectively. After the onset
 of the laser pulse, for $t<t_f$, the spin current changes
non-adiabatically, as displayed in Fig.\ \ref{isas}(a) for different dephasing
angles $\phi$. In this time interval the magnitude of $I_z(t)$ decreases
in time, the variation $I_z(0)-I_z(t)$ being a function of $\phi$. Snapshots
of $I_z(t,\phi)$ recorded at times $t=0,\ 5\hbar/V,\ 8\hbar/V$ are plotted
against the dephasing angle $\phi$ in Fig.\,\ref{isas}(b). We note that
the smallest variation $I_z(0)-I_z(t)$ occurs for $\phi=0$ and the
largest for $\phi=2\pi$.

After the radiation pulse vanishes, i.\ e. for $t>t_f$, the quantum ring
enters the oscillatory regime. Similar to the charge current, the spin current
has an oscillatory behavior with period $T\simeq 2783\hbar/V$ shown in
Fig.\ \ref{isas}(c) where we plot $I_z(t)$.
The amplitude $\Delta I_z$ varies from
$-0.6Va/2$ to $0.6Va/2$ when the angle $\phi$ of the external pulse
is changed from $2\pi/6$ to $10\pi/6$, as seen in Fig.\ \ref{isas}(c).
In addition to the big oscillation with period $T$ and large
amplitude $\Delta I_z\in [-0.6:0.6]Va/2$, the spin current presents
an overlapping pattern of small oscillations, with smaller period
$T'=8\hbar/V$ and smaller amplitude $\Delta' I_z\simeq 0.03Va/2$.
A similar analysis can be done for the spin currents in the
directions $2\theta_\alpha$ and $2\theta_\beta$ (not shown in the
figures).  The small and fast beating oscillation have a lower amplitude
than for $I_z$. In the present example the oscillations of
$I_{2\theta_\beta}$ are slightly larger than those of
$I_{2\theta_\alpha}$ because $V_\beta>V_\alpha$.

\begin{figure}
\includegraphics[width=80mm]{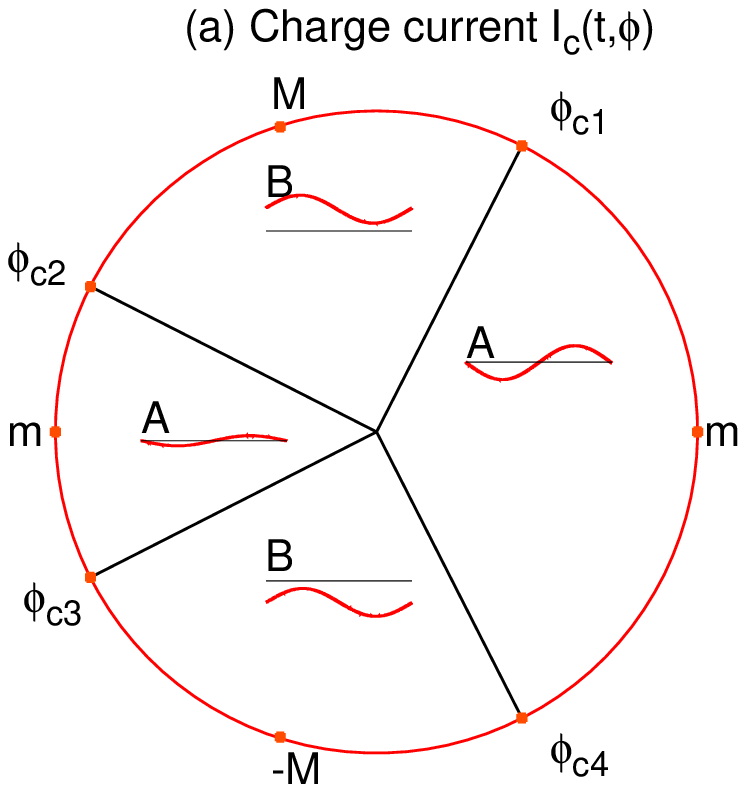}
\includegraphics[width=80mm]{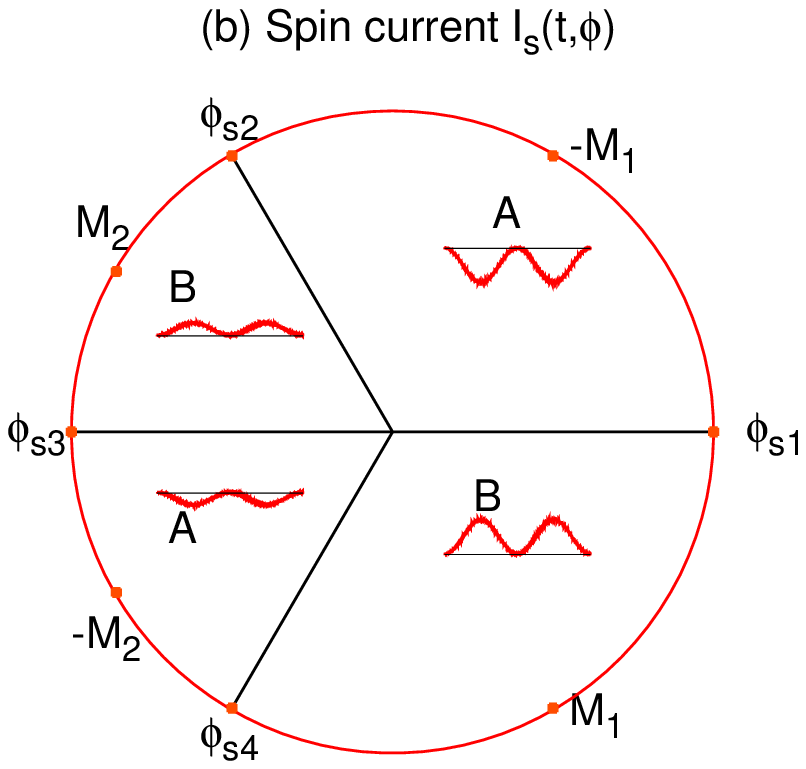}
\caption{(Color online) 
By varying the dephasing angle $\phi$ of the external pulse 
different types of charge and spin current oscillations are induced in the quantum ring.
(a) The charge current has oscillations in time of type A or B. For type A
the average current vanishes for the angles $\phi$ between the dipolar
components of the radiation pulse marked with $m$.  Type B oscillations occur
for $\phi\in [\phi_{c1},\phi_{c2}]$ and $\phi\in [\phi_{c3},\phi_{c4}]$. In this
case the average current has maximum positive and minimum negative amplitudes
for the angles marked with M and -M. At the critical points $\phi_{c1\cdots 4}$
the charge current is constant in time.
(b) The oscillations of the spin current are also classified as type A and B.
For type A the averages current has minimum and negative values at the point
$-M_1$ and $-M_2$. For type B it has maximum positive values at the points
$M_2$ and $M_1$.  The two types of oscillations, A and B, for both $I_c$ and $I_s$,
are in antiphase, respectively.
}
\label{circle}
\end{figure}


The analysis of Fig.\ \ref{isas}(c) indicates
that there are two type of spin current oscillations. For the angle
$\phi=2\pi/6$, $3\pi/6$, $\pi/6$ and $7\pi/6$, the analytic behavior for
the time dependence of spin current is fitted by a function $\cos(2\pi
t/T)$ that defines type A oscillations. For values of the dephasing
angle $\phi=5\pi/6$, $11\pi/6$, $9\pi/6$ and $10\pi/6$ , the analytic
representation is $-\cos(2\pi t/T)$. This defines type B oscillations
that are  $T/2$ dephased from type A.  For certain values of $\phi$ the
amplitude of the broad oscillations of $I_z$, $\Delta I_z$,
become comparable with the amplitude of the fast and small oscillations
of amplitude $\Delta' I_z$.  This happens for $\phi=0$, $\pi$,
$2\pi/3$ and $4\pi/3$ in Fig.\ \ref{isas}(c).  In the later case
the large oscillations with long period $T$ vanish and the remaining
feature is the beating pattern with nodal points already shown in Fig.\
\ref{iz56}.

The amplitude of the big oscillation, $\Delta I_z=I_z(T/2)-I_z(T)$, is
found to be negative for type A and positive for type B oscillations,
which is shown in Fig.\ \ref{isas}(d).  By varying $\phi$ the
spin current switches between A and B type oscillations at four
values of the dephasing angle, denoted as $\phi_{s1,\dots, s4}$.
As shown in Fig.\ \ref{isas}(d) the critical values are 0, $2\pi/3$, $\pi$,
and $4\pi/3$ ($j=0,4,6,8$).  The time average of the spin current,
$\langle I_z\rangle_t=(I_z(T/2)+I_z(T))/2$, is also plotted in
Fig.\ \ref{isas}(c). When $\phi$ changes $\langle I_z\rangle_t$
varies between locally minimum and negative value for $\phi=\phi_{-M1}=2\pi/6$
and for $\phi=\phi_{-M2}=7\pi/6$ reached within type $A$ oscillations,
and locally maximum positive values for $\phi=\phi_{M2}=5\pi/6$ and
$\phi=\phi_{M1}=10\pi/6$ within type B oscillations.  These results
indicate that the type of oscillations performed by the spin current induced
by the two-component radiation pulse, as well as the average spin currents,
can be selected by changing the dephasing angle $\phi$.

The behavior of the spin current along the directions ${\bf
e}_{2\theta_\alpha}$ and ${\bf e}_{2\theta_\beta}$ is qualitatively
similar to the spin current $I_z(t)$, having the same periodicity and
beating structure. The amplitude of the fast oscillations is however
smaller than for $I_z(t)$.

In deriving these results we ignored the inhomogeneity of the electron
distribution around the ring, which is known to appear in the simultaneous presence
of the Rashba and Dresselhaus terms.\cite{sheng,nowak} Underlying this
choice is the fact that the charge deformation in a realistic ring is
small, and even questionable for a narrow two-dimensional ring, unless
placed in an external magnetic field.\cite{nowak} Moreover, when the
electron-electron interaction is considered for more than two electrons,
the charge fluctuation is flattened out due to screening.\cite{daday}
For our one-dimensional ring model the electron density has minima at
polar angles $\pi/4$ and $5\pi/4$ and maxima at $3\pi/4$ and $7\pi/4$.
Therefore, since the circular symmetry is intrinsically broken in the ground state,
a radiation pulse with only one dipolar component would, in principle, be 
sufficient to induce persistent oscillations of the charge and spin 
currents.\cite{nita2}

\section{Summary and conclusions}\label{sec:conc}

We investigate the interference effect generated by the simultaneous
presence of the Rashba and Dresselhaus spin-orbit interactions on charge
and spin currents induced non-adiabatically in a quasi-one-dimensional
ring by a two-component radiation (laser) pulse. Our numerical results are
obtained for a system of few non-interacting electrons through a direct
calculation that involves the exact, time-dependent solution of the density
operator. The main finding of this work is that the oscillatory behavior
of the charge and spin currents is realized at a frequency equal to the difference
between two excited energy states (Bohr frequencies). 

By varying the dephasing angle $\phi$ between the two dipoles of the
external pulse, different types of charge and spin current oscillations
are induced in the ring. The general features are summarized in
Fig.\ \ref{circle}(a) and Fig.\ \ref{circle}(b), respectively.
By changing the internal dephasing angle $\phi$ in the two intervals
$[\phi_{c4},\phi_{c1}]$ and $[\phi_{c2},\phi_{c3}]$ the oscillations of
the charge current in time are qualitatively like  \ $-\sin(2\pi t/T)$,
i.\ e. the oscillations of type A indicated in  Fig.\ \ref{circle}(a). The
average current is zero (i.\ e minimum) at the angles indicated by the
letter "m" in Fig.\ \ref{circle}(a), which are $\phi=0$ and $\phi=\pi$.
For $\phi\in [\phi_{c1},\phi_{c2}]$ and $\phi\in [\phi_{c3},\phi_{c4}]$
the charge current oscillates in time (qualitatively) like the function
$\sin(2\pi t/T)$, which are oscillations of type B, phase shifted with
$T/2$ relatively to the ones of type A. The time average has positive
maximum and negative minimum values at the angles marked with $M$ and $-M$
in Fig.\ \ref{circle}(a).  When the angle $\phi$ has the critical values
$\phi_{c1,\dots,c4}$ a crossover between the two types of oscillations
occurs, and the charge current is constant in time.

The spin current has also such oscillations in time.  When the
angle $\phi$ is in the two intervals $[\phi_{s1},\phi_{s2}]$ and
$[\phi_{s3},\phi_{s4}]$ the induced spin current oscillates in time
as $\cos(2\pi t/T)$, indicated as oscillations of type A in Fig.\
\ref{circle}(b). The time average of the spin current has the minimum
(negative) values at the points marked as $-M_1$ and $-M_2$.  For $\phi\in
[\phi_{s2},\phi_{s3}]$ and $\phi\in [\phi_{s4},\phi_{s1}]$ the spin
current oscillate in time as \ $-\cos(2\pi t/T)$, marked as oscillations
of type B in Fig.\ \ref{circle}(b), and dephased with T/2 relatively to
type A. Their maximum positive values occur at the angles marked
with $M_2$ and $M_1$. These are wide oscillations with long period
$T$ and large amplitude.

In addition, the spin current has also tiny oscillations with a much smaller
period $T'$ and amplitude $\Delta'$.  These high-frequency oscillations
are caused by the nutation of the electron spin between the spin axes
imposed by the R and D couplings.  When the angle $\phi$ is close to the
critical points $\phi_{s1,\dots,s4}$ a crossover occurs between type A
and type B oscillations (or vice versa) of the spin current. For these
values of the angle $\phi$ the amplitude of the wide oscillations of
$I_z$ ($\Delta$) decreases and becomes comparable to the amplitude of
fast oscillations ($\Delta'$). The spin current oscillations reduce to
a beating pattern, while the wide oscillation with long period $T$ vanishes.

After the original excitation disappears the system sustains two persistent 
types of oscillations of the charge and spin currents, which we classified 
as type A and type B.  The two types of oscillations are in antiphase.
Transitions between these modes can be controlled by varying the dephasing 
angle between the two components of the radiation pulse,
an idea with potential applications in the spin-based information technology.

\begin{acknowledgments}

This work was supported by the Icelandic Research Fund, DOE grant number
DE-FG02-04ER46139, and by the Romanian PNCDI2 Research Programmes TE
90/05.10.2011 and Core Programme 45N/2009.

\end{acknowledgments}


\end{document}